\pdfoutput=1

\documentclass[%
preprint,
 amsmath,amssymb,
 aps,
]{revtex4-1}

\usepackage{graphicx}% Include figure files
\usepackage{dcolumn}% Align table columns on decimal point
\usepackage{bm}% bold math

 \usepackage{amssymb}
 \usepackage{float}
\usepackage{times}

\begin{document}
\abovedisplayskip=0.2cm
\abovedisplayshortskip=-0.2cm
\belowdisplayskip=0.1cm
\belowdisplayshortskip=0.1cm

\title{ Principles of Adaptive Sorting Revealed by {\it In Silico} Evolution}

\author{Jean-Beno\^it Lalanne}
\author{Paul Fran\c cois}
\affiliation{Physics Department, McGill University, Montreal, Quebec, Canada H3A 2T8}%

\date{\today}

\begin{abstract}

Many biological networks have to filter out useful information from a vast excess of spurious interactions. We use computational evolution to predict design features of networks processing ligand categorization. The important problem of early immune response is considered as a case-study. Rounds of evolution with different constraints uncover elaborations of the same network motif we name ``adaptive sorting". Corresponding network substructures can be identified in current models of immune recognition. Our work draws a deep analogy between immune recognition and biochemical adaptation.

\begin{description}
\item[PACS numbers]
87.16.Xa, 87.18.Mp, 87.18.Tt, 05.10.-a
\end{description}
\end{abstract}

                             
\maketitle

Information processing in biology often relies on complex out-of-equilibrium physical processes ensuring efficiency \cite{Swain:2002}. The paradigmatic example is kinetic proofreading (KPR), first proposed to explain low spurious base-pair interactions during DNA replication \cite{Hopfield:1974,Ninio:1975}. KPR originated in a context with comparable concentrations of correct and spurious substrates. If the spurious substrate has similar characteristics and is orders of magnitude higher in concentration than the correct one, alternative strategies are needed. 

An important instance of this problem is immune recognition by T cells. T cells constantly scan antigen presenting cells (APC) in their environment, via the binding of their T cell receptors (TCR) to the presented pMHC ligands. T cells perform a sorting process based on interaction with self (non-agonist) or foreign (agonist) ligands at the surface of APCs: if foreign ligands are detected, then the immune response is triggered. Following the ``life-time" dogma \cite{Feinerman:2008b}, one of the main determinants for distinguishing self from foreign is the unbinding time of the pMHC ligand to TCR. Ligands up to a critical binding time of $\tau_c\simeq 3$~s do not elicit response while foreign ligands bound for a longer time ($\tau_f>\tau_c$) do. Self ligands dissociate rapidly (typically for $\tau_s \lesssim 0.1$ s).

\begin{figure}[h!]
\includegraphics[width=0.75\textwidth]{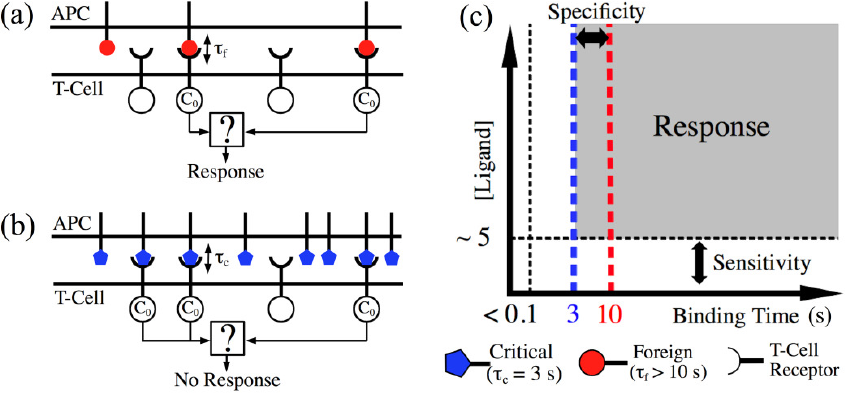}
\caption{\label{fig:network_requirement} (color online) Schematic illustration of the problem setup. (a) Few foreign ligand ($\tau_f > 10$ s) trigger response. (b) Arbitrary large concentrations of critical agonist ($\tau_c = 3$ s) ligands do not trigger response. (c) Idealization of the number of pMHC ligand required to trigger response as a function of pMHC-TCR binding time. Shaded region corresponds to conditions for which the immune response is triggered.}
\end{figure}

The sorting process is extremely sensitive: response is triggered in the presence of minute concentrations of foreign ligands (of the order of 1-10 ligands per cell \cite{Irvine:2002,Feinerman:2008a}). Sorting is specific: although foreign ($\tau_f$) and critical ligands ($\tau_c$) have similar binding times, an arbitrary concentration of critical ligands does not elicit response \cite{AltanBonnet:2005}. These requirements are summarized on  Fig.~\ref{fig:network_requirement}. McKeithan \cite{McKeithan:1995} proposed that T cells harness the amplifying properties of KPR to solve the recognition problem between few foreign ligands and vastly numerous self ligands. However, this model does not account for sharp thresholding required for sensitivity and specificity as noticed in \cite{AltanBonnet:2005}. Other control structures must exist.

We use computational evolution \cite{Francois:2004} to ask the related ``inverse problem" question: how can a network categorize sharply two ligands with similar affinity  irrespective of their concentrations? We exhibit networks performing ligand recognition with the help of a new network module that we name ``adaptive sorting" which we study analytically. We use extensive evolutionary simulations to show how this solution is improved to solve the related recognition problem of parallel sorting of foreign ligands within a sea of self ligands.  We expect the principles presented here to have broader relevance for biological recognition systems where specific signals must be extracted from a high number of weak spurious interactions.

\clearpage

{\it Methods}. -- The algorithm we use to generate biochemical networks is essentially the same as that used in \cite{Francois:2008}, with a biochemical grammar adapted to the specific problem of ligand recognition by (immune) cells. Following the model described in \cite{AltanBonnet:2005}, we limit our possible interaction grammar to phosphorylations or dephosphorylations with rates linear in enzyme concentrations. We assume ligands bind TCR outside the cell, resulting in the activation of the internal part of the receptor (denoted by $C_0$, see Fig.~\ref{fig:network_requirement} (a), (b)).  The algorithm then proceeds to add/remove kinases/phosphatases to evolve cascades of phospho-reactions downstream of $C_0$.  We make the classical hypothesis underlying  KPR models  \cite{McKeithan:1995} that when a ligand dissociates from a receptor, the receptor's internal part gets quickly dephosphorylated. This assumption is consistent with the "kinetic segregation" mechanism \cite{Davis:2006} (see details in \cite{Supplement}). We assume that a single species in the network plays the role of the output of the system and triggers immune response in a binary way via a thresholding mechanism. The nature of the output is under selective pressure and can be changed by the algorithm.

\begin{figure}[h!]
\includegraphics[width=0.75\textwidth]{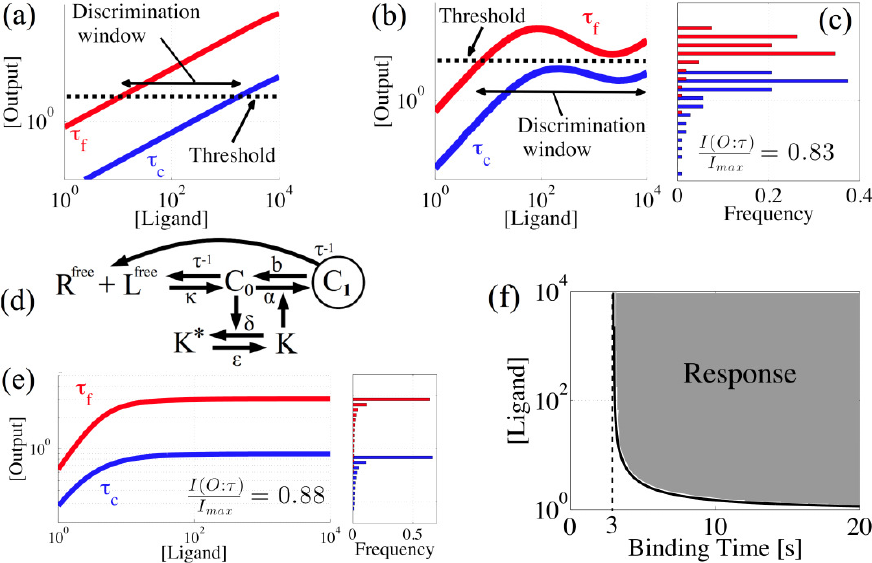}
\caption{\label{fig:adaptive_node} (color online) (a) KPR scheme has discrimination abilities over a limited range of ligand concentration. (b) Output {\it vs} ligand  for $\tau_f = 10$ s and $\tau_c = 3$ s. (c) Histogram of outputs from (b) illustrating effective probability distribution. (d) Adaptive sorting network. Arrows with no specified enzyme represent unregulated reactions. The output is circled.  We keep conventions throughout. (e) Output {\it vs} ligand and histogram of output for adaptive sorting ($\kappa = 10^{-4}$, $R = 10^4$, $\delta = 1$, $\epsilon = 1$, $\alpha = 0.3$, $ b = 0$ and $ K_T = 1$). (f) Minimum ligand concentration triggering response for different binding times for adaptive sorting in (e). Threshold taken to be $\xi( \tau_c)$.}
\end{figure}

The goal here is to discriminate between two kinds of ligands with identical on-rate (denoted by $\kappa$) but different binding times: $\tau_f=10 $ s for foreign ligands and $\tau_c=3 $ s  for critical ligands (we checked that our results do not depend on the specific choice of $\tau_f$ and $\tau_c$ as long as both are of the same order of magnitude). For pure KPR \cite{McKeithan:1995}, the concentration of the output is linear in ligand concentration. Thus, as shown in Fig.~\ref{fig:adaptive_node}~(a), ligands with similar binding times  are distinguished by a thresholding mechanism only over a limited range of concentration, even for a large number of proofreading steps \cite{AltanBonnet:2005}. In contrast, if the steady state output concentration is almost flat in ligand concentration due to some control mechanism, as shown in Fig.~\ref{fig:adaptive_node}  (b), then ligands can be categorized by thresholding nearly irrespective of their concentration.
\vspace{5mm}

To select for networks producing almost flat ligand dependency, we start by sampling logarithmically the range of allowed ligand concentration. Then, steady state outputs are computed for every ligand concentration and binned for the two binding times considered (Fig.~\ref{fig:adaptive_node}~(c) shows the binned outputs corresponding to Fig.~\ref{fig:adaptive_node}~(b)). One then considers the histograms of output for different $\tau$'s  as an effective probability distribution function. A natural measure of performance (``fitness") selecting for networks with behaviour similar to Fig.~\ref{fig:adaptive_node}~(b) is then the mutual information, $\mathcal{I}(O\mbox{;}\tau = \{\tau_c,\tau_f\})$\cite{Tkacik:2011}, between the output value and the dissociation time. A network for which $\mathcal{I}(O\mbox{;}\tau) = \mathcal{I}_{max} (=  1$ bit) has its output distributions for $\tau_f$ and $\tau_c$ disjoint, which is biologically equivalent to a perfect discrimination.  We run our evolutionary simulations with this fitness function for 30 values of ligand concentrations equally spaced on a logarithmic scale in the interval $[1~10^4]$. More details on the evolutionary simulations are given in \cite{Supplement}.

{\it Simple adaptive sorting}. -- We run our simulations with deterministic integration of network equations. Figure~\ref{fig:adaptive_node}~(d) presents a typical network topology we obtain, with the corresponding distribution of outputs on Fig.~\ref{fig:adaptive_node}~(e). Distributions corresponding to the two binding times are clearly separated. In this network, $C_0$ is phosphorylated once into $C_1$ by kinase $K$. $K$ is itself phosphorylated by $C_0$, which makes it inactive. Here, $C_1$ is the output. Calling $R,L$ and $K_T$ the total concentration of receptors, ligands and kinase  respectively equations for this network are
\begin{align}
\dot{C}_0 &= \kappa R^{{\scriptsize \mbox{free}}} L^{{\scriptsize \mbox{free}}} -\left( \alpha K + \tau^{-1} \right) C_0 + b C_1, \label{C0} \\
\dot{C}_1 &= \alpha K C_0 - (\tau^{-1} + b) C_1,  \label{C1}\\
\dot{K} &= -\delta  C_0 K + \epsilon \left(K_T - K \right). \label{K}
\end{align}

$R^{{\scriptsize \mbox{free}}}=R-\sum_{i=0}^{1} C_i$ and $L^{{\scriptsize \mbox{free}}}=L - \sum_{i=0}^{1}  C_i $ are the concentrations of free receptors and ligands. Assuming receptors are in excess ($R^{{\scriptsize \mbox{free}}}\simeq R$), the steady state concentration of output variable $C_1$ can be easily computed and we get $C_{1} = \frac{\xi(\tau) C_0}{C_0+C_*}$ where $\xi(\tau)=\frac{\alpha K_T C_*}{ b+\tau^{-1}},~C_*=\epsilon \delta^{-1}$.

For large $L$, $C_0 \propto L$. In particular, as $C_0 \gg C_*$, $C_1~\simeq~\xi(\tau)$. It is also clear that even for small $L$, $C_1$ will be a pure function of $\tau$ independent from $L$ if $C_*$ small enough.  To discriminate between two ligands with binding times $\tau_1$ and $\tau_2$, one then simply needs to assume response is activated for a $C_1$ threshold value $\theta \in [\xi(\tau_1),\xi(\tau_2)]$. Figure~\ref{fig:adaptive_node} (f) illustrates the range in ligand concentration leading to response with such thresholding (taking $\theta = \xi(\tau_c))$ process for the present network. The network shows both extremely good sensitivity and specificity (compare with Fig.~\ref{fig:network_requirement}~(c)).

This situation is reminiscent of biochemical adaptation, where one variable returns to the same steady state value irrespective of ligand concentration. Indeed, the motif displayed on Fig.~\ref{fig:adaptive_node}~(c) implements an ``incoherent feedforward loop" logic as observed in adaptive systems \cite{Francois:2008,Behar:2007,Ma:2009}: $C_0$ feeds negatively into kinase $K$,  and both $C_0$ and $K$ feed positively into output $C_1$. The overall influence of $C_0$ (and of $L$) is a balance between two opposite effects which cancel out. However, one significant difference from classical adaptation is that the steady state concentration of $C_1$ is now a function of the extra parameter $\tau$, the ligand dissociation time. Discrimination of ligands based on the value of the output becomes possible irrespective of the ligand concentrations.

This process can be generalized to other adaptive networks based on ligand-receptor interaction, as long as one kinetic parameter is ligand specific. For instance, ligand-receptor networks evolved in \cite {Francois:2008} can be modified to have a steady state concentration depending on ligand nature. Call $I$ the input and $R$ the (stable) receptor. If we assume that the complex $C$ resulting from association $I$ and $R$ is washed away with a time constant $\tau _I$ depending on the nature of the input, then the simple adaptive system $\dot{R} = \rho - I R$ and $\dot{C} = IR - C/\tau_I$  stabilizes to a steady state concentration $C=\rho \tau _I $, which depends only on $\tau _I$ irrespective of input value. Schematically, $I$ plays same role as $C_0$ (proportional to ligand concentration in Eq. \ref{C0}) while $R$ plays the role of $K$ (inversely proportional to $C_0$ from Eq. \ref{K} and buffering it in Eq. \ref{C1}) . We believe that this combination of biochemical adaptation with a kinetic parameter dependency to perform decision can potentially be observed in a wide variety of biochemical networks. We subsequently call it {\it adaptive sorting}.

\begin{figure}[h!]
\includegraphics[width=0.75 \textwidth]{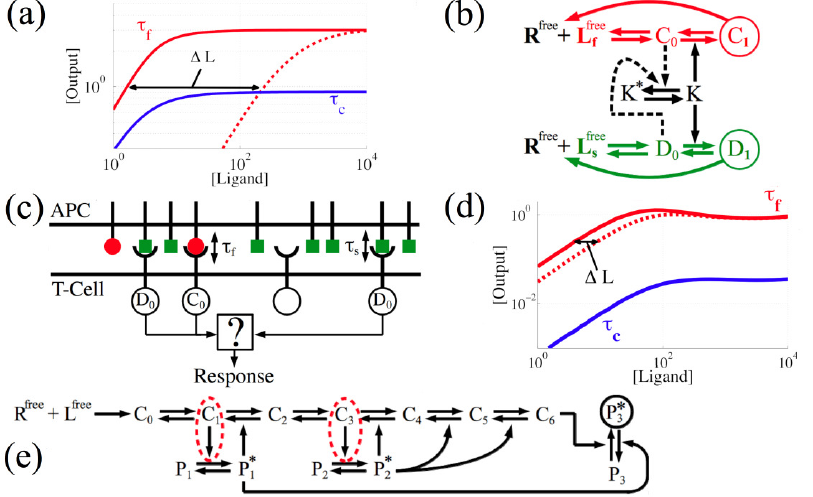}
\caption{\label{fig:downstream_activation} (color online)  (a) Effect of self ligands on the adaptive sorting module from  Fig.~\ref{fig:adaptive_node}~(e), taking $[C_N]+[D_N]$ as an output. Full lines: $L_s= 0$. Dashed line: $L_s = 10^4$. Note the catastrophic effect of self ligands on sensitivity (quantified through $\Delta L$). We compare $\tau_f$ with $L_S > 0$ to $\tau_c$ with $L_s = 0$ as a worst case scenario. (b) Coupling (dashed arrows) between two different types of ligands through kinase $K$ for adaptive sorting. (c) Schematic illustration of new constraint of parallel sorting. Squares represent self ligands ($\tau_s = 0.05$ s). (d) Example of evolved output {\it vs.} ligand relathionship with $L_s = 0$ (full) and $L_s = 10^5$ (dashed). Loss in sensitivity is now small. (e) Schematic of network corresponding to (d). Complexes $C_i$'s are understood to decay to $R^{{\scriptsize \mbox{free}}}$ and $L^{{\scriptsize \mbox{free}}}$ (same convention in Fig.~\ref{fig:summary}). Parameters are given in \cite{Supplement}. }
\end{figure}

{\it Parallel adaptive sorting}. -- Adaptive sorting by itself is efficient to discriminate independently critical from foreign, but its performance is degraded when cells are exposed {\it at the same time} to foreign ligands (concentration $L_f$) and a huge excess of self ligands (concentration $L_s$), as illustrated in Fig.~\ref{fig:downstream_activation}~(a). This phenomenon is not specific to the immune system and is called antagonism \cite{AltanBonnet:2005}.  The reason is that the two different kinds of ligands are coupled through the common kinase used in the feedforward motif (dashed arrows in Fig.~\ref{fig:downstream_activation}~(b)). Precisely, denoting the complexes arising from the binding of foreign and self ligands by $C_i$ and $D_i$ respectively, the total output concentration is 
\begin{align}\label{eqn:C_1_with_self}
C_{1} +D_{1}\simeq C_{1}= \frac{\xi(\tau_f) C_0}{C_0+D_0+C_*},
\end{align}
which still tends to $\xi(\tau_f)$ at large $L_f$. We can neglect $D_1$ in the output because $\xi(\tau)\propto \tau$ and so $\xi(\tau_s) \ll \xi(\tau_f)$. To reach the adaptive regime, we now have the requirement that $C_0 \gg D_0$. For large $L_s$, $D_0 \gg D_1$ and we have that  $D_0 + D_1\simeq D_0 \approx \kappa R \tau_s (1+\kappa R \tau_s)^{-1} L_s$. Similarly, $C_0\simeq \kappa R \tau_f (1+\kappa R \tau_f)^{-1} L_f$. Thus $C_1 \simeq \xi(\tau_f)$ for
\begin{align}\label{eqn:condition_adaptation_as}
L_f \gg \left(\frac{1 + \kappa R\tau_f}{1 + \kappa R\tau_s}\right)  \left(\frac{\tau_s}{\tau_f}\right) L_s \sim \kappa R \tau_s  L_s
\end{align}

With $\kappa R \tau_f \gg1$, $L_s \sim 10^5$ and $\kappa R \tau_s \sim 0.1$, self ligands thus annihilate the sensitivity of the simple adaptive sorting motif.

To solve this problem, we rerun evolutionary simulations with the constraint that discrimination between $\tau_f$ and $\tau_c$ should happen even in the massive presence of self ligands ($\tau_s=0.05$~s), as sketched in Fig.~\ref{fig:downstream_activation}~(c). A representative result of this computational evolution is presented in Fig.~\ref{fig:downstream_activation}~(d) and (e) for output and network topology respectively. The networks found look very similar to adaptive sorting, except that the incoherent feedforward module is sometimes implemented via activation of a phosphatase, instead of de-activation of a kinase \footnote{It can be shown that regulation via the phosphatase indeed requires at least two phosphorylation steps, explaining why it is less probable to evolve compared to the motif of Fig.~\ref{fig:adaptive_node} (d) when no other constraint is imposed.}. A full cascade of KPR also evolves. Notably, in all working networks there is an important difference with the previous case:  activation of the enzyme in the adaptive sorting module is rewired downstream the first step of the KPR cascade (dashed circles in Fig.~\ref{fig:downstream_activation} (e)).

\vspace{5mm}

\begin{figure}[H]
\centering
\includegraphics[width=0.75 \textwidth]{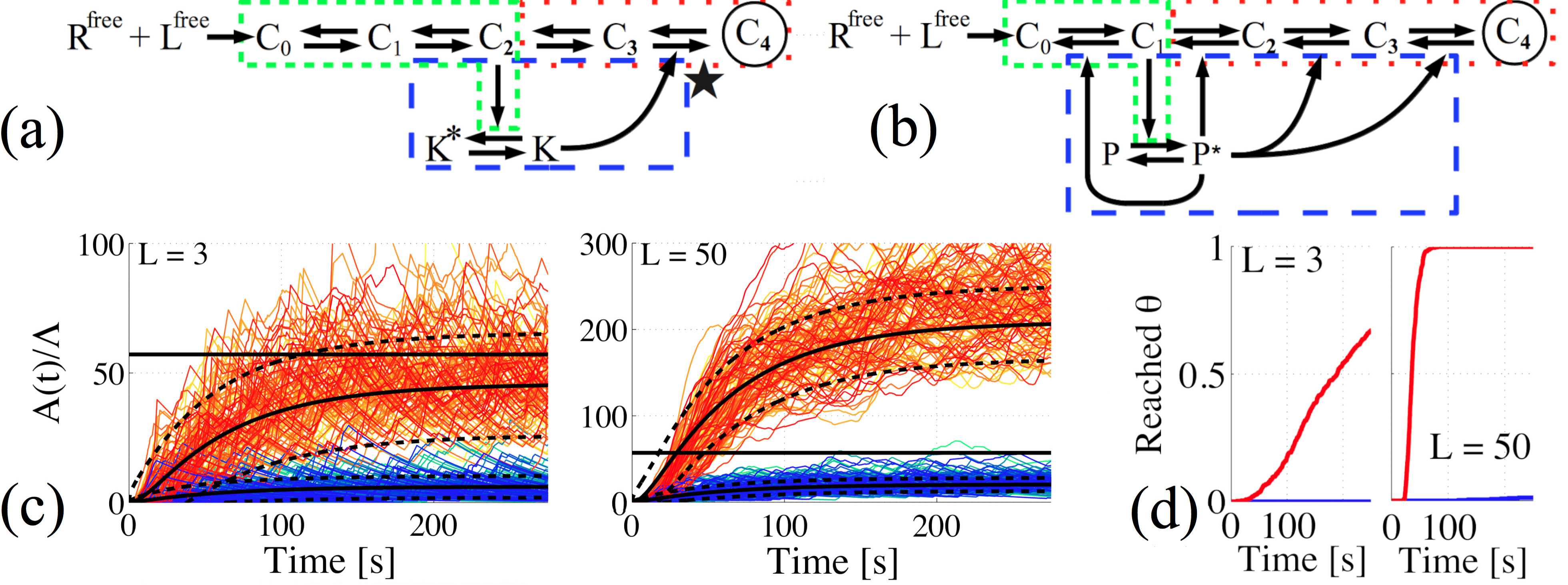}
\caption{\label{fig:summary} (color online) (a) Final network with categorization properties in presence of large concentrations of spurious substrates. The parallel (long dash), adaptive (fine dash) and KPR (dotted) modules are identified. Star indicates the specific phosphorylation in adaptive sorting. (b) Network for immune recognition with corresponding features, from \cite{AltanBonnet:2005, Francois:2013}. Adaptive sorting is achieved via the activation of non specific phosphatase $P^*$, assumed to be SHP-1. (c) Sample trajectories of $A$ for different ligand concentrations $L$. Warm color for $\tau_f$ and cold colors for $\tau_c$. Threshold $\theta$ is identified by an horizontal line. Black curves are the analytic expressions $\langle A(t) \rangle \pm \sigma_{A(t)}$ \cite{Supplement}. (d) Fraction of trajectories having reached threshold for $N = 4$, $m = 2$, $\kappa = 10^{-4}$, $R = 10^4$, $\delta = 1$, $\epsilon = 0.5$, $\phi = 0.3$, $\alpha = 0.0003$, $ b = 0$, $ K_T = 1000$, $L_s = 0$. }
\end{figure}

This can be fully understood analytically by considering an idealized network such as the one in Fig.~\ref{fig:summary}~(a) which is compared to the actual network implicated in immune response \cite{AltanBonnet:2005,Francois:2013} in Fig.~\ref{fig:summary}~(b). Our idealization consists in an adaptive sorting module with upstream and downstream steps of KPR ($N$ steps in total, adaptive module activated by complex $m$, $m+2\leq N$,  Fig.~\ref{fig:summary}~(a)). In such networks, assuming no dephosphorylation down the cascade ($b = 0$), the output takes the form \cite{Supplement}
\begin{align}
C_{N} +D_{N}\simeq C_N = \frac{\xi'(\tau_f) C_0}{ C_m + D_m + C_*(1+\alpha K_T \tau_f)},
\end{align}
where $C_m = \gamma_f^m C_0$ and $D_m = \gamma_s^m D_0$, with $\gamma_i = \phi \tau_i (1+\phi\tau_i)^{-1}$. $\phi$ denotes the default (unregulated) phosphorylation rate in the cascade. $\xi'(\tau)$ is a function of $\tau$, and like before $\xi'(\tau_s)\ll \xi'(\tau_f)$ so that we can neglect the contribution of $D_N$ in the output. Even in the presence of many self ligands $L_s$, we clearly have an output independent of $L_f$  for $C_0 \gg \gamma_f^{-m} \gamma_s^{m} D_0$  ( $m=0$ is simple adaptive sorting). Since $\phi \sim \tau_f^{-1}$ for a sensitive network \cite{Supplement}, $\gamma_s \gamma_f^{-1}$ is small, thus any $m>1$ makes $\gamma_f^{-m} \gamma_s^{m}$ even smaller.   This is in essence a weak proofreading process upstream the cascade ensuring that $C_m\gg D_m$ so that the adaptive sorting module is only triggered by foreign ligands. As for simple adaptive sorting, we have that $C_0 \propto L_f$ and $D_0 \propto L_s$ although the prefactors differ \cite{Supplement}. In the end, $C_N$ is a pure function of $\tau_f$ for 
 \begin{align}
L_f \gg \left(\frac{1 + \kappa R\tau_f}{1 + \kappa R\tau_s}\right) \left(\frac{\gamma_s}{\gamma_f} \right)^{m+1}  L_s \label{eqn:antagonism}
\end{align}
so that the r.h.s is small compared to Eq.~\ref{eqn:condition_adaptation_as} for $m>0$. Self influence is consequently almost abolished.

\vspace{5mm}

It must be emphasized that the more complex solutions displayed in Fig.~\ref{fig:downstream_activation}~(e) and Fig.~\ref{fig:summary}~(a) require more than one kinase or phosphatase: generic enzymes are shared by most of the proofreading steps, while a specific enzyme accounts for the adaptive sorting module (star in Fig.~\ref{fig:summary} (a)). This is of biological importance since it is not clear that biochemistry would allow fine-tuned specificity to a single step in the cascade. Interestingly, alternative solutions also evolve where kinases and phosphatases are not specific to a given proofreading step \cite{Supplement}. For these networks, discrimination is still possible, but loss of biochemical specificity degrades the adaptive properties. Instead, one observes a non-monotonic behaviour, flattened out over the range of input ligand considered, so that adaptation is only approximated. For a complete analytic and experimental study of such a case, see \cite{Francois:2013}.

{\it Dealing with low numbers of molecules }- Immune cells perform efficient sorting of different ligand types for as little as $\sim 10$ foreign ligands. A low number of molecules is potentially problematic because adaptive sorting shows a trade-off between specificity and sensitivity. In the simpler scheme ( Fig. \ref{fig:adaptive_node}~(c)) , perfect adaptation for $L\rightarrow 0$  occurs if $C_* \rightarrow 0 $, but output in the adaptive regime is $C_1 = \xi(\tau) \propto C_*\rightarrow 0$ so that discrimination becomes impossible. Increasing $N$ actually softens the constraint: KPR steps downstream the adaptive module (Fig. \ref{fig:summary}) add a geometric dependency in $\tau$ to $C_N$ , so that $C_N$ can have a strong dependency in $\tau$ (specificity) even for low $C_*$ (sensitivity) \cite{Supplement}.

 Another potential problem comes from fluctuations at low ligands. In the immune context, fully phosphorylated tails of receptors (corresponding to $C_N$ in our model) themselves slowly phosphorylate abundant ($>10^4$ molecules) downstream targets such as ZAP70 and ERK, which saturates and triggers response only after a couple of minutes  \cite{AltanBonnet:2005}. Following \cite{Francois:2013}, we check that coarse-graining this downstream cascade into a slow variable solves the fluctuation problem. We pose a variable $A$ obeying $\dot{A} = \Lambda C_N - T^{-1} A$. $A$ is a proxy for the abundant targets and can therefore be realistically assumed to be deterministic as long as $\Lambda$ is large, so that the only $A$ stochasticity comes from $C_N$.  We assume thresholding is then made on the deterministic value of $A$, leading to a binary {\it irreversible} decision \cite{Lipniacki:2008}. We take $T = 60$ s, as the response of T-cells occurs on the order of minutes \cite{AltanBonnet:2005}.

Simulations of this process using Gillespie algorithm are presented in Fig. \ref{fig:summary}~(c) and (d), with samples of trajectories and fraction of activated cells as a function of time. Results are in very good agreement with a simple linear noise approximation on $C_N$(see details and assumptions in \cite{Supplement}- in particular, fluctuations of $A$ decay as $T^{-1/2}$). Ligands at $\tau_c$ essentially never cross the threshold for the considered time window, while for ligands at $\tau_f$,  almost all cells eventually respond for $L_f>5$ (this number goes to $\lesssim 10$ in presence of self ligands \cite{Supplement}). Finally, the model's half population response time (Fig.~\ref{fig:summary} (d)) is consistent with experiments \cite{Supplement, AltanBonnet:2005, Francois:2013} and decreases down to less than one minute as $L_f$ increases. So, although we cannot exclude that other noise-resistance mechanism are possible \cite{Wylie:2007},  adaptive sorting coupled to a slow downstream cascade has discrimination capabilities compatible with experimental data.

Many biological systems have to filter out specific useful information from a vast excess of spurious interactions. We have evolved {\it in silico} networks categorizing ligands with very close biochemical properties  irrespective of their concentrations. We have discovered a new functional unit, the adaptive sorting module, which can be rewired to solve a parallel sorting problem. Our final model is summarized in Fig.~\ref{fig:summary}~(a), along with corresponding network features of the immune system Fig.~\ref{fig:summary}~(b) \cite{Francois:2013}. Strikingly, the network of the immune system shares many similarities with our final solution. In our framework, immune recognition corresponds to an optimal solution with non-specific enzymes. We expect adaptive sorting to manifest itself through non linear (or even non monotonic) dependency of response on input concentration. This is observed in a wide range of networks, for instance in endocrine signalling \cite{Vandenberg:2012}, but remains mechanistically unexplained. Adaptive sorting could lie at the core of such signalling processes as well as others.

We thank Eric Siggia, Massimo Vergassola, Guillaume Voisinne and Gr\'egoire Altan-Bonnet for useful discussions. JBL is supported by NSERC, PF by NSERC and HFSP.

\clearpage

\begin{centering}
{\Large Supplementary Information}\\
\end{centering}
\vspace{-4mm}
\begin{small}
\tableofcontents
\end{small}

\newpage

\section{Grammar of Allowed Chemical Reactions}\label{sec:details_evo}

The starting point of all our evolutionary simulations is the minimal network shown below.
\begin{figure}[h!]
\begin{center}
\includegraphics[width=0.15\textwidth]{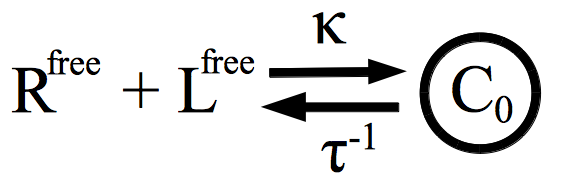}
\end{center}
\end{figure}
\vspace{-6mm}

With corresponding equation
\[
\dot{C}_0 = \kappa \left(R - C_0\right) \left(L - C_0\right) - \tau^{-1} C_0,
\]
where $L$ and $R$ are the total concentrations of ligand and receptor respectively. $R^{{\scriptsize \mbox{free}}} = R - C_0$ is the free receptor concentrations and $L^{{\scriptsize \mbox{free}}} = L-C_0$ is the free ligand concentration. The circle denotes that $C_0$ is the output of the network. Biologically, we consider $C_0$ to be the "activated" intracellular section of the TCR bound to a pMHC.

The grammar of the network allows for the complex $C_0$ to be phosphorylated:
\begin{figure}[h!]
\begin{center}
\includegraphics[width=0.25\textwidth]{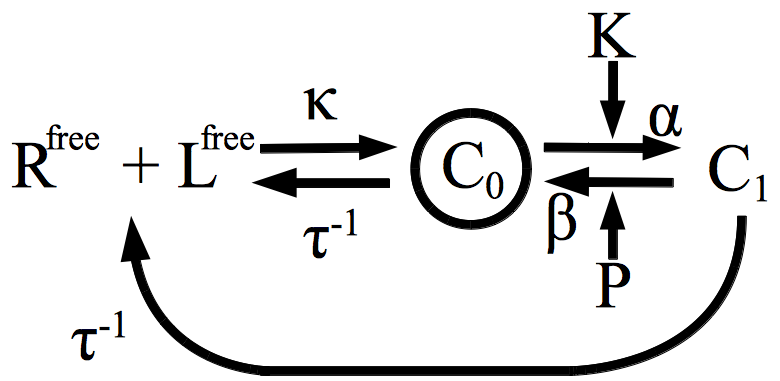}
\end{center}
\end{figure}
\vspace{-5mm}

We model the phosphorylations and dephosphoarylations in the simplest possible scheme. The equations for this network are then
\begin{eqnarray}
\dot{C}_0 &=& \kappa \left(R_T - C_0 - C_1\right) \left(L_T - C_0 - C_1\right) - \left(\tau^{-1} + \alpha K\right) C_0 + \beta P C_1, \nonumber \\
\dot{C}_1 &=& \alpha K C_0 - \left(\beta P + \tau^{-1} \right) C_1. \nonumber 
\end{eqnarray}

Here, $C_0$ gets phosphorylated to $C_1$ by kinase $K$. Each phosphorylation is always associated with a dephosphorylation (to avoid irreversible reactions). In this network, the dephosphorylation is catalysed by phosphatase $P$. Observe also that $C_1$ is taken to decay to free ligands and receptors with rate $\tau^{-1}$. This reaction amounts to a kinetic proofreading step. The decay of $C_1$ directly to receptors and ligands is equivalent to demanding fast dephosphorylation of the internal section of the TCR relative to the binding time $\tau$ (the importance of this assumption is discussed in more details in Sec.~\ref{sec:non_infinite_unbound_dephosphorylation}). Note that the output is left under selective pressure (i.e. the output tag is not carried to $C_1$ automatically). The grammar allows for this reaction (phosphorylation of $C_n$) to occur an unlimited number of times. 

We assume that complexes in the cascade (the $C$'s) are kinases. This means that $C_n$ can phosphorylate any kinase or phosphatase, except other complexes $C_i$. To have the reaction grammar as unbiased as possible, we take the catalytic activity of any kinase (or phosphatase) to initially be shared with its phosphorylated counterpart (e.g., if initially $K$ phosphorylates $C_0$ and $K$ gets phosphorylated to $K^*$, then $K^*$ also phosphorylates $C_0$). The algorithm allows for the removal of interactions in subsequent generations if that proves advantageous in terms of the chosen fitness. To take a concrete example, if in the previous reaction $C_0$ phosphorylates $K$ with rate $\delta$, the network becomes as below.

\begin{figure}[h!]
\begin{center}
\includegraphics[width=0.25\textwidth]{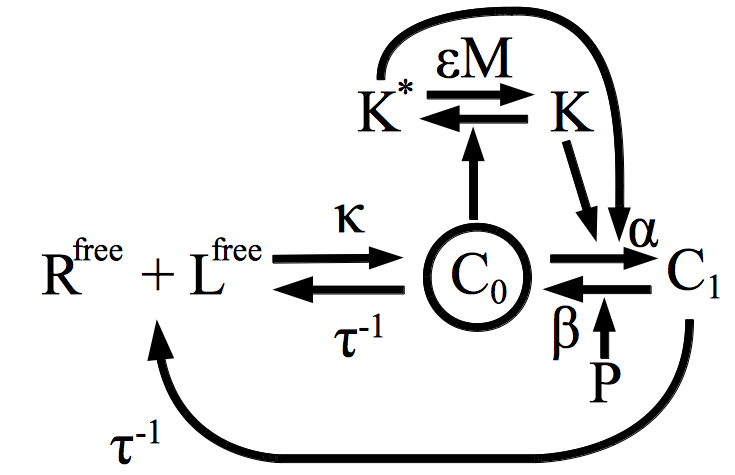}
\end{center}
\end{figure}

With corresponding equations
\begin{eqnarray}
\dot{C}_0 &=& \kappa \left(R_T - C_0 - C_1\right) \left(L_T - C_0 - C_1\right) - \left\{\tau^{-1} + \alpha (K + K^* )\right\} C_0 + \beta P C_1, \nonumber \\
\dot{C}_1 &=& \alpha (K + K^*) C_0 - \left(\beta P + \tau^{-1} \right) C_1, \nonumber \\
\dot{K} &=& -\delta C_0 K + \epsilon M (K - K_T). \nonumber
\end{eqnarray}
Observe how $K + K^* = K_T$, and not just $K$, enters the equation of $C_0$ and $C_1$. This inheritance of catalytic activity also holds for complexes in the kinetic proofreading cascade (the $C$'s).

As mentionned before, reactions can also be removed by the algorithm. For example, the reaction of $K^*$ with $C_0$ could be removed. The output tag can also shift between different species. For instance, the output could become $C_1$. One then ends up with network

\begin{figure}[h!]
\begin{center}
\includegraphics[width=0.25\textwidth]{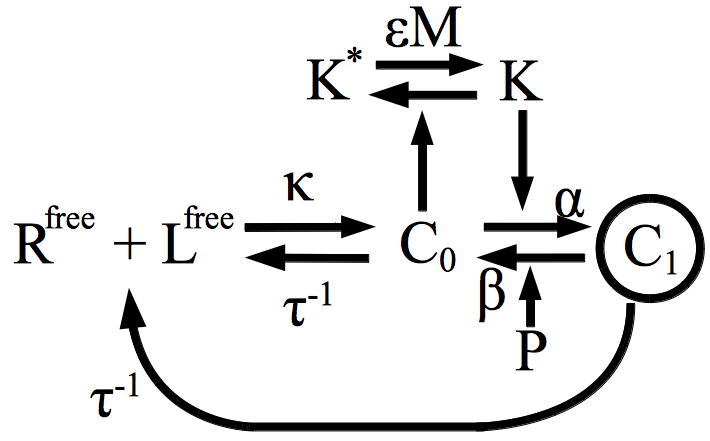}
\end{center}
\end{figure}

In which case the equations read
\begin{eqnarray}
\dot{C}_0 &=& \kappa \left(R_T - C_0 - C_1\right) \left(L_T - C_0 - C_1\right) - \left(\tau^{-1} + \alpha K \right) C_0 + \beta P C_1, \nonumber \\
\dot{C}_1 &=& \alpha K C_0 - \left(\beta P + \tau^{-1} \right) C_1, \nonumber \\
\dot{K} &=& -\delta C_0 K + \epsilon M (K - K_T). \nonumber
\end{eqnarray}

This is just the adaptive sorting module.

It must be made explicit that our grammar assumes that the catalysis have fast kinetics. This means that an enzyme concentration is unaffected by its catalytic activities. Moreover, the strongest non-linearities are quadratic. This assumption is mainly motivated by simplicity. Even with these simple ingredients, solutions with interesting features are found.

Finally, the grammar allows for more than one kinase to phosphorylate a given reaction and similarly for phosphatases. Also, there is a possibility to add kinases and phosphatase (we typically start evolution with two kinases and phosphatases present). In addition to having dynamic topologies in the algorithm, all kinetic parameters as well as concentrations of enzymes not on the $C$ cascade can be modified within predetermined range.

Initial conditions for the integration of the networks' equations are taken to be as biologically realistic as possible. Before cell-cell contact, there clearly no bound TCR and pMHC. We therefore take $C_i = 0 ~\forall i$ at $t = 0$ (the antigen presenting cells contacts the T cell at time 0).

When performing evolution with cells exposed simultaneously to self and foreign ligands, we make coupling between self and foreign complexes as shown in the case of simple adaptive sorting in Fig.~3~(b) of the paper. If the output is a member of the cascade of complexes, then it is taken to be the sum of complexes arising form self as well as foreign ligands. We run our simulations for 30 values of ligand concentrations equally spaced on a logarithmic scale in the interval $[1~10^4]$.

\section{Derivation of Asymptotic Output Concentrations}\label{sec:solutions}

\subsection{Simple Adaptive Sorting}\label{sec:plain_vanilla}
We consider the following system of differential equations (shown schematically in Fig.~2~(d) of paper)
\begin{eqnarray}\label{eqn:simple_solution}
\dot{C}_0 &=& \kappa \left(R-\sum_{i = 0}^1 \left[C_i + D_i\right] \right) \left(L_f - C_0 - C_1 \right) -\left( \alpha K + \tau_f^{-1} \right) C_0 + b C_1, \nonumber \\
\dot{C}_1 &=& \alpha K C_0 - (\tau_f^{-1} + b) C_1, \nonumber \\
\dot{D}_0 &=& \kappa \left(R-\sum_{i = 0}^1 \left[C_i + D_i\right] \right) \left(L_s - D_0 - D_1 \right) -\left( \alpha K + \tau_s^{-1} \right) D_0 + b D_1, \nonumber \\
\dot{D}_1 &=& \alpha K D_0 - (\tau_s^{-1} + b) D_1, \nonumber \\
\dot{K} &=& -\delta \left( C_0 + D_0\right) K + \epsilon \left(K_T - K \right).
\end{eqnarray}

The $C$'s and the $D$'s are respectively the agonist (foreign or critical) and the non-agonist (self) complexes. The output is $C_1 + D_1$. The following tables summarize the meaning of each variable.
\begin{center}
\begin{tabular}{c|c}
Symbol & Chemical Species (Concentration)\\
\hline
\hline
$C_n$ & Agonist complex phosphorylated n times \\
$D_n$ & Non-agonist complex phosphorylated n times \\
$L_s$ & Self ligands, binding time $\tau_s = 0.05$ s (total)\\
$L_c$ & Critical agonist ligands, binding time $\tau_c = 3$ s (total)\\
$L_f$ & Foreign ligands, binding time $\tau_f = 10$ s (total)\\
$R$ & Receptor (total)\\
$K$ &  Kinase \\
$K_T$ & Kinase (total)\\
\end{tabular}
\end{center}
\vspace{3mm}

\begin{center}
\begin{tabular}{c|c|c}
Symbol & Kinetic Parameter & Order\\
\hline
\hline
$\alpha$ & Complex phosphorylation rate & 2\\
$b$ & Complex dephosphorylation rate & 1\\
$\tau_s$ & Self complex binding time & -\\
$\tau_c$ & Critical agonist complex binding time& -\\
$\tau_f$ & Foreign complex binding time& -\\
$ \delta$ & Kinase phosphorylation rate & 2\\
$\epsilon$ & Kinase dephosphorylation rate& 1\\

\end{tabular}
\end{center}

We suppose that the receptors are largely in excess, i.e., $\left(R-\sum_{i = 0}^1 \left[C_i + D_i\right] \right) \approx R$ We are interested in the steady-state concentration (see Sec.~\ref{sec:transient}). We first consider the case where the are no self ligands, $L_s = 0$. The case of two ligand types is treated in sec.~\ref{sec:two_ligands_simple}. Our goal is to determine the behaviour of $C_1$ (steady-state) as function of $L_f$. Unless otherwise specified, we take $\tau_f = 10$ s, $\tau_c = 3$ s and $\tau_s = 0.05$ s.

Adding the two equations for $C_0$ and $C_1$ in equation (\ref{eqn:simple_solution}), we get
\begin{equation}\label{eqn:sum_D}
C_0 + C_1 = \left(\frac{\kappa R \tau_f}{\kappa R \tau_f + 1}\right) L_f.
\end{equation}

$C_0 + C_1$ equals to the concentration of bound receptors. We can thus assess our assumption that the free receptors outnumber greatly the bound receptors. Substituting typical concentrations $L_f$ and assuming $\kappa R \sim 1$, we do get that $C_0 + C_1 \ll R$. The condition is stretched a bit for self ligands (which are more numerous), but the analytical results derived are nonetheless in excellent agreement with numerical results for all but unrealistically large ligand concentrations.

Now, we have at steady state 
\begin{equation}\label{eqn:D_1}
C_1 = \frac{\alpha K C_0}{\tau_f^{-1} + b}~~~\mbox{and}~~~K = \frac{K_T}{1 + \left(\frac{\delta}{\epsilon}\right) C_0}~~~\Rightarrow ~~~C_1 = \frac{\alpha K_T C_0}{\left(\tau_f^{-1} + b\right) \left(1 + \left(\frac{\delta}{\epsilon}\right) C_0\right). }
\end{equation}

Observe how $C_1 \propto K C_0$ and $K \propto {C_0}^{-1}$ at large $C_0$. This shows right away that $C_1$ will be independent of $C_0$ and thus of $L_f$ at large enough ligand concentrations. One can substitute equation~(\ref{eqn:D_1}) in equation~(\ref{eqn:sum_D}) to obtain

\begin{equation}\label{eqn:restriction_L}
C_0 \left( 1 + \frac{\alpha K_T}{\left(\tau_f^{-1} + b\right) \left(1 + \left(\frac{\delta}{\epsilon}\right) C_0\right) } \right) = \left(\frac{\kappa R \tau_f}{\kappa R \tau_f + 1}\right) L_f.
\end{equation}

This is a quadratic equation for $C_0$. It has only one positive solution, whose exact form is not important, but which reduces for large $L_f$ to

\[
C_0 \rightarrow \left(\frac{\kappa R \tau_f}{\kappa R \tau_f + 1}\right) L_f - \frac{C_{*} \alpha K_T \tau_f}{1 + b \tau_f}~~~~~\mbox{for}~L_f~\mbox{large.}
\]

We defined $C_{*} \equiv \epsilon \delta^{-1}$. For large $L_f$ we then have, considering equation~(\ref{eqn:D_1}):
\begin{equation}\label{eqn:D_1_simple}
C_1 \rightarrow \frac{C_{*} \alpha K_T \tau_f}{1 + b \tau_f}.
\end{equation}

$C_1$ monotonically approaches this limit from below. %In the case where self ligands are present, we see that since $\tau_s \ll \tau_f$, the contribution in output due to self ligands will be negligible.

The last important quantity to solve for is the scale of concentration of $L_f$ for which $C_1$ is close to its asymptotic value. A natural measure is the ligand concentration for which the output reaches one half of its asymptotic value. We see from equation~(\ref{eqn:D_1}) that this occurs when $C_0 = C_{*}$. Solving for $L_f$ in such case yields
\[
L_{f,1/2} = C_{*} \left(\frac{\kappa R \tau_f + 1}{\kappa R \tau_f}\right) \left( 1 + \frac{\alpha K_T \tau_f}{2\left(1 + b\tau_f\right)}\right).
\]

More generally, one can show that $L_{f,\Lambda}$ defined through $C_{1}(L_{f} = L_{f,\Lambda}) = \Lambda~\cdot~C_1(L_{f}~\rightarrow~\infty)$ equals, from equation~(\ref{eqn:D_1}) and (\ref{eqn:restriction_L}) (and taking $b=0$ to compare with the parallel sorting case)

\begin{equation}\label{eqn:L_delta_N_1}
L_{f,\Lambda} = C_{*} \left( \frac{\kappa R \tau_f + 1}{\kappa R \tau_f} \right) \left( \frac{\Lambda}{1-\Lambda} \right) \left(1 + [1 - \Lambda] \alpha K_T \tau_f\right).
\end{equation}

\subsection{Parallel Adaptive Sorting}\label{sec:general_calculation}

Below is the general case with upstream and downstream kinetic proofreading of the adaptive module. $N$ is the total number of steps in the signalling cascade. $m$ denotes the complex activating the adaptive node. The symbols have the same meaning as in Sec.~\ref{sec:plain_vanilla}. The only restriction is that 0 $\leq$ m $\leq$ N-2. 

\begin{eqnarray}\label{eqn:system}
\dot{C}_0 &=& \kappa \left(R-\sum_{i = 0}^N \left[C_i + D_i\right] \right) \left(L_f-\sum_{i = 0}^N C_i \right) -\left( \phi + \tau_f^{-1}\right) C_0, \nonumber \\
\dot{C}_n &=& \phi \left( C_{n-1} - C_n \right) - \tau_f^{-1} C_n ~~~~~~~\mbox{for } 1\leq n \leq N-2, \nonumber \\
\dot{C}_{N-1} &=& \phi C_{N-2} - \left( \alpha K + \tau_f^{-1} \right) C_{N-1}, \nonumber \\
\dot{C}_{N} &=& \alpha K C_{N-1} - \tau_f^{-1} C_N, \nonumber \\
\dot{D}_0 &=& \kappa \left(R-\sum_{i = 0}^N \left[C_i + D_i\right] \right) \left(L_s-\sum_{i = 0}^N D_i \right) -\left( \phi + \tau_s^{-1}\right) D_0, \nonumber \\
\dot{D}_n &=& \phi \left( D_{n-1} - D_n \right) - \tau_s^{-1} D_n ~~~~~~~\mbox{for } 1\leq n \leq N-2, \nonumber \\
\dot{D}_{N-1} &=& \phi D_{N-2} - \left( \alpha K + \tau_s^{-1} \right) D_{N-1}, \nonumber \\
\dot{D}_{N} &=& \alpha K D_{N-1} - \tau_s^{-1} D_N, \nonumber \\
\dot{K} &=& -\delta \left( C_m + D_m\right) K + \epsilon \left(K_T - K \right).
\end{eqnarray}

\begin{figure}[h!]
\centering
\includegraphics[width=0.75\textwidth]{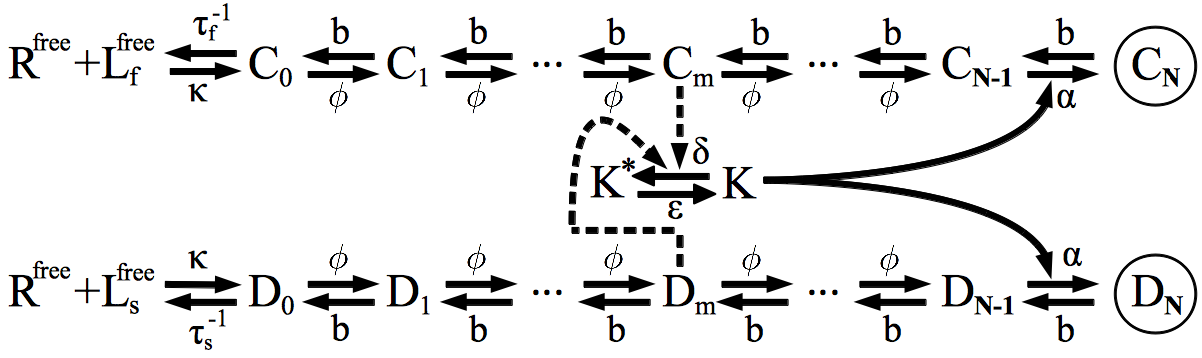}
\caption{\label{fig:schematic_generalized_adaptive_sorting} Schematic representation of parallel adaptive sorting. Foreign complexes ($C_i, i>0$) decay to $R^{{\scriptsize \mbox{free}}}$ and $L^{{\scriptsize \mbox{free}}}_f$ with rate $\tau_f^{-1}$ (not shown). Self complexes ($D_i, i > 0$) decay to $R^{{\scriptsize \mbox{free}}}$ and $L^{{\scriptsize \mbox{free}}}_s$ with rate $\tau_s^{-1}$ (not shown). The output here is $C_N + D_N$ (circled). All arrows with no specified enzyme are unregulated (fixed rate).}
\end{figure}

The schematic representation of this network is shown in Fig.~\ref{fig:schematic_generalized_adaptive_sorting}. 

\vspace{-2mm}
The only new variable is $\phi$, which is the (unregulated) phosphorylation rate down the cascade. In order to make analytic progress, we take the dephosphorylation rate $b = 0$. The qualitative features of the solutions do not depend on that assumption (see Sec.~\ref{sec:non_zero_dephosphorylation} for a discussion).

Observe that the only form of coupling between the self and foreign is through receptor sequestration (the term $R-\sum_{i = 0}^N \left[C_i + D_i\right]$) and phosphorylation of the kinase (shown as dashed arrows in Fig.~\ref{fig:schematic_generalized_adaptive_sorting}). As before, we take $R^{{\scriptsize \mbox{free}}} = R-\sum_{i = 0}^N \left[C_i + D_i\right] \sim R$ in the above equations. We consider uniquely the steady state. Summing the equations for the $C_i$, one obtains
\begin{equation}\label{eqn:constraint_sum}
\sum_{i = 0}^N C_i = \left(\frac{\kappa R \tau_f}{\kappa R \tau_f + 1}\right) L_f.
\end{equation}

Now, note that
\[
C_n = \left(\frac{\phi \tau_f}{\phi \tau_f +1} \right) C_{n-1}~~~~~\mbox{for } 1\leq n \leq N-2.
\]

To alleviate the notation slightly, define $\gamma_f \equiv \ \phi \tau_f(\phi \tau_f +1)^{-1} $. One can then repeat the above to obtain $C_n$ in terms of $C_0$, that is
\begin{equation}\label{eqn:C_n}
C_n = \gamma_f^n C_0~~~~~\mbox{for } 1\leq n \leq N-2.
\end{equation}

One can also easily obtain that from the original system of equations (\ref{eqn:system})

\begin{equation}\label{eqn:C_N-1}
C_{N-1} = \left( \frac{\phi \tau_f}{\alpha K \tau_f +1} \right) C_{N-2} = \left( \frac{\phi \tau_f}{\alpha K \tau_f +1} \right) \gamma_f^{N-2} C_0,
\end{equation}
\noindent and
\begin{equation}\label{eqn:C_N}
C_{N} = \alpha K \tau_f C_{N-1} = \left( \frac{\alpha K \tau_f }{\alpha K \tau_f +1} \right) \phi \tau_f \gamma_f^{N-2} C_0.
\end{equation}

We can substitute equations~(\ref{eqn:C_n}), (\ref{eqn:C_N-1}) and (\ref{eqn:C_N}) into equation~(\ref{eqn:constraint_sum}). Interestingly, the factor of $K$ (containing the coupling between the two types of ligands) cancels from this equation. We have
\[
\sum_{i = 0}^N C_i = C_0 \left\{\frac{1-\gamma_f^{N-1}}{1-\gamma_f} +  \phi \tau_f \gamma_f^{N-2} \right\} =\left(\frac{\kappa R \tau_f}{\kappa R \tau_f + 1}\right) L_f.
\]

From the definition of $\gamma$, we can further simplify this to
\begin{equation}\label{eqn:C_0}
C_0 =\left(\frac{\kappa R \tau_f}{\kappa R \tau_f + 1}\right) \left(\frac{L_f}{1 + \tau_f \phi}\right).
\end{equation}

The above derivation did not depend the fact that we were specifically considering foreign ligands. The above expression is then also valid for $D_0$ (with $L_f \rightarrow L_s$ and $\tau_f \rightarrow \tau_s$).

We are interested in the output $C_N + D_N$. To solve for it, we must determine the steady-state value of $K$. From our system of equation, we read
\begin{equation}\label{eqn:kinase_K}
K = \frac{K_T}{1 + C_*^{-1} \left( C_m + D_m \right) } = \frac{K_T}{1 + C_*^{-1} \left( \gamma_f^m C_0 + \gamma_s^m D_0 \right) }. 
\end{equation}

As before, $C_{*} = \epsilon \delta^{-1}$. Since $D_0$ and $C_0$ are both known functions of $L_s$ and $L_f$ respectively, then so is $K$. We can rewrite equation~(\ref{eqn:C_N}) as
\begin{equation}\label{eqn:final_result}
C_{N} = \left( \frac{C_* \alpha K_T \phi \tau_f^2 \gamma_f^{N-2}  }{C_*(1+ \alpha K_T \tau_f) + \gamma_f^m C_0 + \gamma_s^m D_0} \right) C_0.
\end{equation}

This is equation (7) from the paper. The equation for $D_N$ is the same (foreign parameters changed to self ones). $D_0$ and $C_0$ are known, so the above represent a closed solution. 

Suppose we consider only one type of ligands (e.g. foreign). Then, since $C_0 \propto L_f$, we have for large $L_f$ that $C_N$ tends to
\begin{equation}\label{eqn:asymptotic_D_N}
C_{N} \rightarrow  \left(\alpha K_T C_{*}  \phi \right) \tau_f ^2 \left( \frac{\phi \tau_f}{1+\phi \tau_f} \right)^{N-2-m}~~~~~\mbox{for $L_f$ large}.
\end{equation}

Observe that $C_N$ is monotonic in $C_0$. The above thus constitute an upper bound on the output. We see then that $C_N$ is at least quadratic in $\tau$. It follows that in the presence of self ligands ($L_s > 0$) the contribution of output from self ligands is then truly negligible, since $\tau_s \ll \tau_f$.

It is desirable to determine the scale of $L_f$ at which this asymptotic limit is attained. We can compute the concentration in foreign ligands, denoted $L_{f}(\Lambda)$, required to attain a fraction $\Lambda$ of the asymptotic output. One can arrive from equation (\ref{eqn:C_0}) and (\ref{eqn:final_result}) at
\begin{equation}\label{eqn:L_delta_parallel}
L_f(\Lambda) =  C_{*} \left( \frac{\Lambda}{1-\Lambda}\right)  \frac{(1+\phi \tau_f)^{m+1}}{\left( \phi \tau_f \right)^m} \left(\frac{\kappa R \tau_f + 1}{\kappa R \tau_f} \right) \left\{ 1+ \alpha K_T \tau_f + C_{*}^{-1} \gamma_s^m D_0 \right\}.
\end{equation}

Notice that the result does not depend on N, the total number of steps in the cascade. Since we want a small $L_f(\Lambda)$, $\phi$ cannot be arbitrarily small or large as $L_f(\Lambda)$ grows as $\phi \rightarrow \infty$ and as $\phi \rightarrow 0$. 

The contribution to $L_f(\Lambda)$ due to $L_s > 0$ will be small given that $\gamma_s \gamma_f^{-1}$ is small (see Sec.~\ref{sec:shift_L} for a discussion). $\gamma_s \gamma_f^{-1}$ is a monotonically increasing function of $\phi$, with value $\tau_s \tau_f^{-1}$ for $\phi = 0$ and tending to $1$ for $\phi \rightarrow \infty$. At $\phi = \tau_f^{-1}$, we have that $\gamma_s \gamma_f^{-1} < 2\tau_s \tau_f^{-1} \ll 1$ already.

The intrinsic contribution (that present if $L_s = 0$) has a non-monotonic behaviour in $\phi$. If we plausibly assume $\alpha K_T = \phi$ (taking all steps in cascade to have same default rate of phosphorylation), then, for a given $C_*$, the minimum of $L_f(\Lambda)$ occurs for $\phi =\frac{1}{2} m \tau_f^{-1}$. Taking $\phi \rightarrow 0$ could decrease self antagonism by a factor of two. However, this would be at the detriment of the intrinsic sensitivity, which would end up dominating at very small $\phi$. $\phi \sim \tau_f^{-1}$ sets the appropriate scale for a maximal sensitivity (optimizing intrinsic sensitivity and mitigating self antagonism).

\subsection{Effects of Non-Zero Cascade Dephosphorylation Rate}\label{sec:non_zero_dephosphorylation}

The analytic calculation for the parallel adaptive sorting module was performed assuming a vanishing dephosphorylation rate (denoted by $b$) down the signalling cascade. In order to better understand the effect of $b >0$, we can first consider the simple adaptive sorting module. This case can be solved with $b>0$. From equation (\ref{eqn:D_1_simple}), we have

\[
\frac{O(\tau_f,b)}{O(\tau_c,b)} = \left(\frac{\tau_f}{\tau_c}\right)\left(\frac{1 + b\tau_c}{1 + b\tau_f}\right)~~~~\mbox{where}~O(\tau,b) = \lim_{L\rightarrow \infty} C_1(L,\tau,b).
\]

We see that the ratio in output arising from foreign and critical ligands is a monotonically decreasing function of $b$. This is intuitively clear: in the limit of large $b$, the dominant opposing force to the production of the output is no longer unbinding of the ligand-receptor complex (which depends on the parameter of interest, $\tau$), but rather the backward reaction rate down the cascade. It is thus expected that the output concentration strongly depends on $b$ and only weakly on $\tau$ in the limit where $b\tau_f \gg 1$.

These observations for simple adaptive sorting actually apply to the parallel sorting module. For $b\tau_f \gg 1$, the output due to different binding time are indistinguishable and all specificity is lost. Figure~\ref{fig:non_zero_dephosphorylation} below illustrates the results of numerical integration for a chosen set of parameters. Qualitatively, the behaviour is the same as simple adaptive sorting.

\begin{figure}[h!]
\begin{center}
\includegraphics[width=0.85\textwidth]{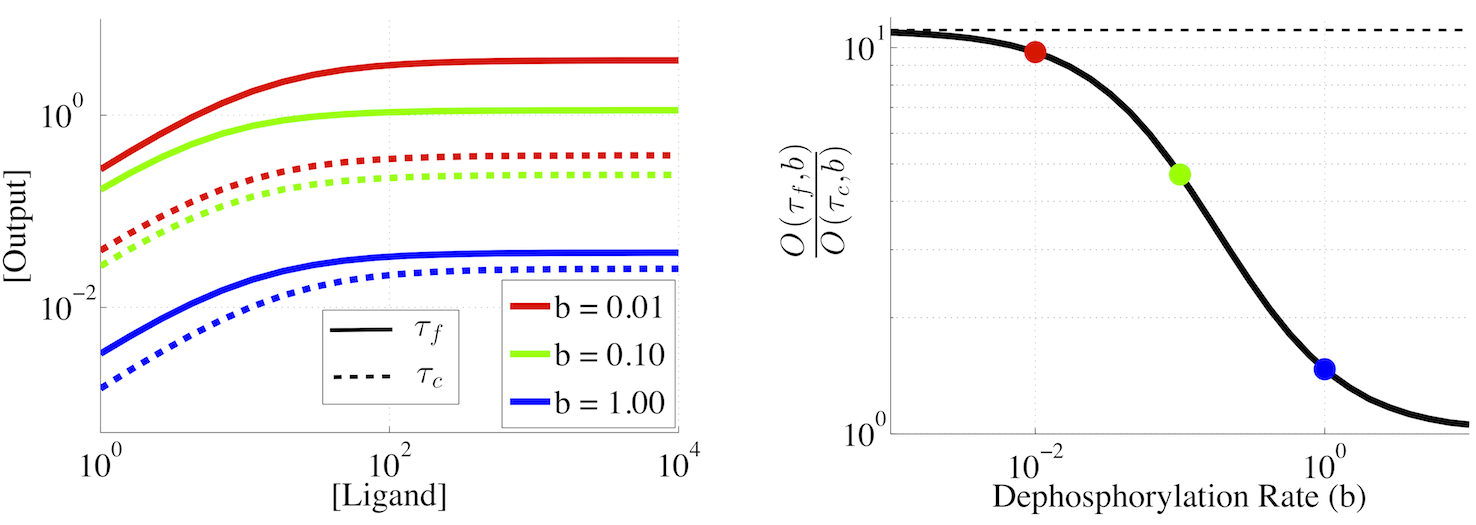}
\caption{\label{fig:non_zero_dephosphorylation} Left panel shows output versus ligand for $\tau_f$ and $\tau_c$ (full and dashed lines respectively) for different values of the dephosphorylation rate $b$. Right panel shows the ratio in foreign and critical output (large ligand concentration asymptotic value) as a function of dephosphorylation rate evaluated numerically (solid line). $b$ values from left panel are identified as dots. The dashed line is the $b=0$ limit (from equation (\ref{eqn:asymptotic_D_N})). Note how dephosphorylation degrades specificity: at large dephosphorylation rate down the cascade, the output from foreign and critical ligands tend to the same value. Parameters: $N = 4$, $m = 2$, $L_s = 0$, $R = 3\times10^4$, $\kappa = 10^{-4}$, $\phi = \alpha = 0.3$, $K_T = 1$, $\delta = 1$ and $\epsilon = 0.5$.}
\end{center}
\end{figure}

As discussed in Sec.~\ref{sec:non_infinite_unbound_dephosphorylation}, assuming a low dephosphorylation rate for bound pMHC-TCR and a high dephosphorylation rate for unbound pMHC-TCR is consistent with the actual biological system of interest even there might appear to exist tension between these requirements.

\section{Discussion of the Specificity/Sensitivity Trade-off (Intrinsic)}\label{sec:specificity_sensitivity_discussion}
As we detail below, the biochemical networks considered in the previous section cannot be arbitrarily sensitive (respond at very low concentration of ligand) and be specific (have very different output concentrations for not very different ligands). The limits derived concern the ``intrinsic" properties of the networks (i.e. in absence of self ligands). A discussion of the effects of self ligands on sensitivity is presented in the next section.

\subsection{Simple Adaptive Sorting}\label{sec:tradeoff_specificity_sensitivity}

The threshold in output at which the system responds is an important variable of our model.  To derive the minimal possible sensitivity, it is desirable to take the threshold to be as small as possible while still retaining the self-consistency of our framework.

Critical agonist ligands are defined to be the ligands for which an arbitrarily high concentration of ligands does not trigger response. One must then demand that the threshold be above the maximum output value that can be attained by the critical ligands ($\tau_s \sim 3$ s):
\[
\mbox{Threshold} > \mbox{Asymptotic output critical critical ligands} = \frac{C_{*} \alpha K_T \tau_c}{1 + b \tau_c}.
\]

This is a hard bound needed for the internal consistency of our model. If the threshold were placed lower, it would imply that critical ligands could trigger a false positive. Note that in practice, in the presence of fluctuations, the threshold must be placed higher to avoid having fluctuations cause response for critical ligands. This point is treated in detail in Sec.~\ref{sec:noise}. Here, we are concerned with strict lower bounds on the sensitivity and in particular in Sec.~\ref{sec:shift_L} on the minimal effect of self ligands on this sensitivity.

We can use equation~(\ref{eqn:L_delta_N_1}) to compute the foreign ligand concentration needed to reach that minimal threshold. To do so, we must take (taking $b = 0$ gives us a strict minimum)
\[
\Lambda = \frac{C_1(L_c \rightarrow \infty)}{C_1(L_f \rightarrow \infty)} = \frac{\tau_c}{\tau_f}.
\]
Substituting this $\Lambda$ in equation~(\ref{eqn:L_delta_N_1}) yields
\begin{equation}\label{eqn:L_thresh}
L_{min} = C_{*} \left(\frac{1 + \kappa R \tau_f}{\kappa R \tau_f} \right) \left\{ \frac{\tau_c}{\tau_f - \tau_c} + \alpha K_T \tau_c \right\}
\end{equation}

$L_{min}$ specifies the minimum sensitivity of the model: ligands with $\tau = \tau_f$ will not trigger response at lower concentrations. $L_{min}$ is optimally as small as possible, since we would like the response to be triggered by very few foreign ligands.

Concerning specificity, note that from equation~(\ref{eqn:D_1_simple}) we must have
\begin{equation}\label{eqn:dO_dtau}
\frac{\mbox{d}C_1(L_f\rightarrow \infty)}{\mbox{d}\tau} \leq C_{*} \alpha K_T.
\end{equation}

The above is a measure in our model of the specificity, i.e., of how fast the asymptotic output concentration varies with the binding time. Optimally, we would like this quantity as large as possible. Small difference in ligand types (characterised by $\tau$) could then be mapped to large differences in output concentrations.

Looking at equations~(\ref{eqn:L_thresh}) and (\ref{eqn:dO_dtau}), we see however that the requirements of arbitrarily low $L_{min}$ and high $\mbox{d}C_1/\mbox{d}\tau$ cannot be met. In effect, we have the constraint (eliminating $C_*$ from the equations)
\[
L_{min} \geq \left(\frac{1 + \kappa R \tau_f}{\kappa R \tau_f} \right)  \left\{ \frac{\tau_c}{\alpha K_T (\tau_f - \tau_c)} + \tau_c \right\} \left(\frac{\mbox{d}C_1}{\mbox{d}\tau}\right) > \tau_c \left(\frac{\mbox{d}C_1}{\mbox{d}\tau}\right).
\]

The prefactor in front of $\mbox{d}C_1/\mbox{d}\tau$ quantifies the intrinsic trade-off between specificity and sensitivity.

\subsection{Parallel Adaptive Sorting}\label{sec:specificity_sensitivity_relief}

It is possible to apply the same idea in the case of adaptive sorting with upstream and downstream kinetic proofreading (Sec.~\ref{sec:general_calculation}). The notion of minimal threshold presented in the previous sub-section holds (i.e., that the threshold must be larger than the asymptotic output concentration corresponding to critical agonist ligands). In particular, using equation~(\ref{eqn:L_delta_parallel}) with the appropriate $\Lambda$, one can obtain $L_{min} $ in the more general case.

As before, it is possible to eliminate one parameter ($C_{*}$) in favour of $\mbox{d}C_N/\mbox{d}\tau$ in the expression for $L_{min}$ to obtain a bound of the form
\[
L_{min} \geq \Gamma_{N,m} (\phi, \alpha K_T) ~\left(\frac{\mbox{d}C_N}{\mbox{d}\tau}\right)_{\tau_M}, ~~~\mbox{where}
\]

\begin{scriptsize}
\[
\Gamma_{N,m}(\phi,\alpha K_T) = \left(\frac{\tau_M}{1 + \phi \tau_M}\right) \left(\frac{1 + \phi \tau_M}{\phi \tau_M} \right)^{N-m} \left(\frac{\tau_c^2 \tau_f^{-2}}{N -m + 2 \phi \tau_M}\right) \left( \frac{\tau_f^{-1}+\phi}{\tau_c^{-1} + \phi}\right)^{N-m-2} \left(\frac{1 + \phi \tau_f}{\phi \tau_f} \right)^{m} \left( \frac{1 + \kappa R\tau_f}{\kappa R \tau_f} \right)
\]
\end{scriptsize}

\begin{figure}[h!]
\begin{center}
\includegraphics[width=0.9\textwidth]{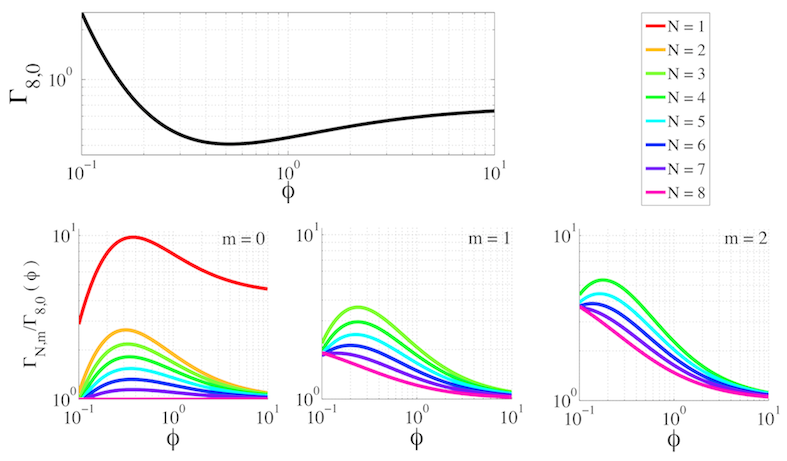}
\caption{\label{fig:gamma_N} Top panel shows $\Gamma_{8,0}$ as a function of $\phi$. We explicitly assume $\alpha K_T = \phi$. Lower panels show $\Gamma_{N,m}$ normalized by $\Gamma_{8,0}$. From left to right: $m = 0, 1$ and 2. Observe how, at fixed $m$, $\Gamma_{N,m}$ decreases with increasing $N$. Moreover, notice that at fixed $N$, $\Gamma_{N,m}$ increases with $m$. $\kappa R = 3$ in the above.}
\end{center}
\end{figure}

Here, $\tau_M= 0.5 (\tau_f + \tau_c)$. Introducing $\tau_M$ is needed since in the general case $\mbox{d}C_N/\mbox{d}\tau$ depends on $\tau$. To get a representative value of $\mbox{d}C_N/\mbox{d}\tau$, we take $\tau$ to be the mean of $\tau_f$ and $\tau_c$. $\Gamma_{N,m}$ (defined only for $m + 2 \leq N$) quantifies the severity of the trade-off between the intrinsic specificity and sensitivity. Clearly, the specific form of the function is not critical. What is important is that at a fixed $m$, increasing $N$ decreases $\Gamma_{N,m}$. Similarly, at a fixed $N$, increasing $m$ increases $\Gamma_{N,m}$. Precisely, one can show that $\Gamma_{N,m} = (1 + \phi^{-1} \tau_c^{-1})^m~\Gamma_{N,0} > \Gamma_{N,0}$. Such behaviour can be traced back to the form of the solution in the general case, equation~(\ref {eqn:asymptotic_D_N}), which contains a factor $\left\{ \phi \tau (1 + \phi \tau)^{-1} \right\}^{N-m-2}$. Hence, larger $N-m$ implies stronger dependency on $\tau$, making it easier to evade the intrinsic specificity-sensitivity trade-off. Fig.~\ref{fig:gamma_N} illustrates these trends. In the figure, we take $\alpha K_T = \phi$ to reduce the number of parameters. This simply means that the phosphorylation catalyzed by kinase $K$ is at the same default rate than the other ones in the cascade. To be clear: small $\Gamma$ implies less trade-off (better).

The $m=0$ case is important as it shows how $N>1$ drastically relaxes the trade-off. For $m>0$, the improvement is still substantial.

\section{Effect of Self Ligands on Sensitivity}\label{sec:shift_L}

\begin{figure}[h!]
\begin{center}
\includegraphics[width=0.93\textwidth]{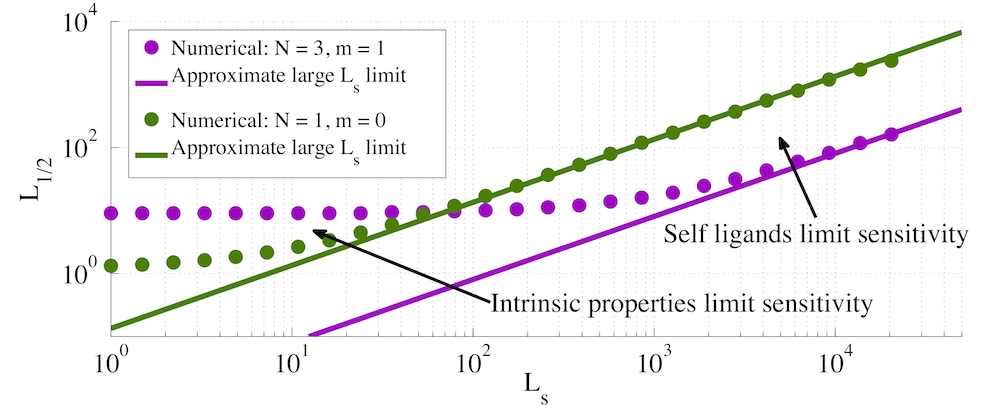}
\caption{\label{fig:comparison_L_half} Ligand concentration required to reach half of maximum output concentration. Results for both simple adaptive sorting (green) and parallel adaptive sorting (purple, example with $N = 3, m =1$) are shown. Parameters (same for both simple and parallel adaptive sorting): $R = 3\times10^4$, $\kappa = 10^{-4}$, $\phi = \alpha = 0.25$, $b = 0$, $K_T = 1$, $\delta = 1$ and $\epsilon = 0.5$. Dots are obtained from numerical integration of the network equations. Lines are the quantities appearing in equations (5) (green) and (7) (purple) in the paper. Note that at low $L_s$ values, the ``intrinsic" sensitivity dominates (see Sec.~\ref{sec:specificity_sensitivity_discussion}).}
\end{center}
\end{figure}

As heuristically shown in the paper, self ligands have deleterious effects on the sensitivity of the adaptive sorting networks. Fig.~\ref{fig:comparison_L_half} compares heuristic conditions (5) and (7) from the paper to numerical solutions for $L_{1/2}$ (foreign ligand concentration required to reach half maximum of asymptotic output).

In the following sub-sections, we derive mathematically rigorous bounds for the loss in sensitivity due to self ligands in both the simple and parallel adaptive sorting modules. This is essentially an elaboration of the results of equation (5) and (7) from the paper.

\subsection{Simple Adaptive Sorting} \label{sec:two_ligands_simple}

In the presence of two types of ligands, trying to solve for the steady-state of equation~(\ref{eqn:simple_solution}) yields a quartic equation. The exact solution yields no insight. Instead, we derive a strict lower bounds on the shift in the minimum ligand value required for response (defined in Sec.~\ref{sec:specificity_sensitivity_discussion}) due to the presence of self ligands. 

In order to obtain a lower bound on the loss of sensitivity due to self ligands, we first obtain an upper bound on $C_1$. To do so we need a lower bound on $D_0$. This is because $D_0$ appears in the equation for $C_1$ (through the kinase which couples the two types of ligands).

$D_0$ is a monotonically decreasing function of the kinase concentration. Setting $K$ to its maximum value $K_T$ yields

\[
D_0 \geq \left(\frac{\kappa R \tau_s}{\kappa R \tau_s + 1}\right) \frac{L_s}{\left( 1 + \frac{\alpha K_T}{\tau_s^{-1} + b} \right)}  > \left(\frac{\kappa R \tau_s}{\kappa R \tau_s + 1}\right) \frac{L_s}{\left( 1 + \alpha K_T \tau_s \right)} \equiv D_{0,low}.
\]

The lower bound in $D_0$ then gives us an upper bound on the value of K in the steady-state

\[
K \leq \frac{K_T}{1 + C_*^{-1} \left(C_0 + D_{0,low}\right)}.
\]

This finally gives us an upper bound on the concentration of $C_1$, satisfied for all concentrations of self ligands
\[
C_1 = \frac{\alpha K C_0}{\tau_f^{-1} + b}  \leq \frac{K_T}{1 + C_*^{-1} \left(C_0 + D_{0,low}\right)}  \left(\frac{\alpha C_0}{\tau_f^{-1} + b}\right).
\]

We now see that our bound on $C_1$ is a monotonically increasing function of $C_0$. It is possible to bound $C_0$ above
\[
C_0 \leq C_0 + C_1 = \left(\frac{\kappa R \tau_f }{\kappa R \tau_f + 1} \right) L_f \equiv C_{0,up}.
\]

This gives us our final rigorous upper bound on $C_1$
\[
C_1 \leq \frac{K_T}{1 + C_*^{-1} \left(C_{0,up} + D_{0,low}\right)}  \left(\frac{\alpha C_{0,up}}{\tau_f^{-1} + b}\right) \equiv C_{1,up}.
\]

Note that the presence of self ligand will contribute through $D_1$ to the output concentration (taken to be $C_1 + D_1$). Because of the smallness of $\tau_f^{-1} \tau_s$,  this contribution is however negligible.

We can obtain via $C_{1,up}$ the minimum foreign ligand concentration needed to reach the minimal threshold value in the presence of self ligands. This allows us to obtain a lower bound on the loss of sensitivity in presence of self ligands  $\Delta L_{min}(L_s)\equiv L_{min}(L_s) - L_{min}(L_s = 0)$. We get
\begin{equation}\label{eqn:bound_delta_L_as}
\Delta L_{min}(L_s) > \left\{\frac{\tau_{c}}{\tau_f - \tau_{c}} \right\} \left(\frac{1 + \kappa R \tau_f}{1 + \kappa R \tau_s} \right)  \left( \frac{\tau_s}{\tau_f} \right) \frac{L_s}{1 + \alpha K_T \tau_s}.
\end{equation}

The term in $\{ \}$ depends on the specific choice of threshold. The second term is slightly smaller than the scale stated in the paper (equation (5)) because of $\alpha K_T \tau_s$ (which was neglected in our heuristic argument). Numerical integration show that this bound is indeed satisfied (see Fig.~\ref{fig:delta_L_numeric}). It is known that $L_s \sim 10^5$ on the surface of antigen presenting cells. Hence, our simple but rigorous bounds tells us right away that if the adaptive node is activated by $C_0$, then $\Delta L_{min}(L_s) \gtrsim 2000$ where data has response around $L \sim 5$. This means that in the presence of multiple self ligands, the ability of this simple model to respond at low foreign ligand concentrations is annihilated.

\subsection{Parallel Adaptive Sorting}\label{sec:general_self}

Interestingly, the more complicated case (Sec.~\ref{sec:general_calculation}) is solvable exactly. In particular, equation~(\ref{eqn:L_delta_parallel}) allows us to compute $\Delta L_{min}(L_s)$ defined in the previous sub-section. Indeed, $\Delta L_{min}(L_s)$ is the contribution coming from the self ligands ($D_0$) in equation~(\ref{eqn:L_delta_parallel}), so

\[
\Delta L_{min}(L_s) =  \left( \frac{\Delta}{1-\Delta}\right)  \frac{(1+\phi \tau_f)^{m+1}}{\left( \phi \tau_f \right)^m} \left(\frac{\kappa R \tau_f + 1}{\kappa R \tau_f} \right)  \gamma_s^m D_0.
\]

Substituting for the expression for $D_0$ in terms of $L_s$ (equation~(\ref{eqn:C_0})), and using $\Lambda$ consistent with the minimal possible threshold, one arrives at (neglecting terms of order $\Delta^2$ since $\Delta \sim 0.1$)

\begin{equation}\label{eqn:delta_L_general_as}
\Delta L_{min}(L_s) = \left\{\left( \frac{\tau_c}{\tau_f}\right)^{2} \left( \frac{\tau_f^{-1}+\phi}{\tau_c^{-1} + \phi}\right)^{N-m-2} \right\} \left(\frac{1 + \kappa R \tau_f}{1 + \kappa R \tau_s } \right) \left( \frac{\tau_f^{-1}+\phi}{\tau_s^{-1} + \phi}\right)^{m+1} L_s.
\end{equation}

The term in $\{ \}$ is specific to our choice of threshold, but the second term corresponds to the scale given in the paper (equation (7)). The relevant term is the last factor before $L_s$. Since $\phi \sim \tau_f^{-1}$, this is a small parameter. We see thus that increasing $m$ (i.e. moving the activation of the adaptive complex downstream) strongly decreases the ability of the non-agonist ligand to desensitize response. Comparison of equation~(\ref{eqn:delta_L_general_as}) to numerical evaluation is illustrated in Fig.~\ref{fig:delta_L_numeric} below.

\begin{figure}[h!]
\centering
\includegraphics[width=0.8\textwidth]{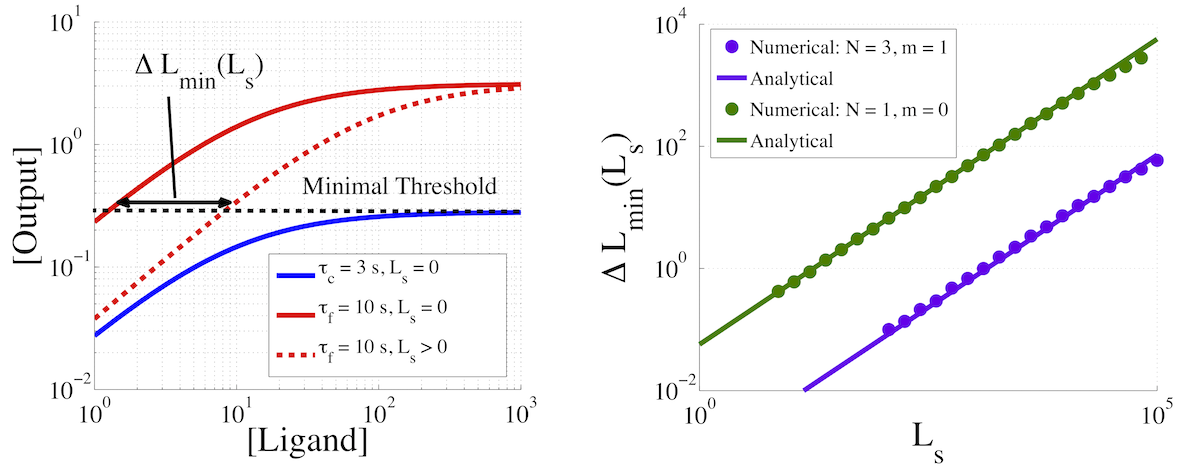}
\caption{\label{fig:delta_L_numeric} Left panel shows schematically the definition of $\Delta L_{min}$ as a quantifier of loss of sensitivity due to self ligands. Full lines are for $L_s = 0$ and the dashed line for $L_{s} > 0$. Right panel shows numerical calculation of $\Delta L$ together with analytical results of equations (\ref{eqn:bound_delta_L_as}) (green) and (\ref{eqn:delta_L_general_as}) (purple) for simple adaptive sorting and parallel adapative sorting respectively (calculations with $N = 3$ and $m = 1$). Parameters are the same as in Fig.~\ref{fig:comparison_L_half}. There is slight disagreement at large $L_s$ due to receptor saturation.}
\end{figure}

It is not possible here to quantify a trade-off between specificity and sensitivity as cleanly as for the intrinsic case (Sec.~\ref{sec:specificity_sensitivity_discussion}). Such a trade-off between minimizing the self antagonism and maximizing specificity is nevertheless present. To see this, note that the only parameter that we could change in equation~(\ref{eqn:delta_L_general_as}) is $\phi$. $\Delta L_{min} (L_s)$ is a monotonically decreasing function of $\phi$. Our result for $C_N$ (equation~(\ref{eqn:asymptotic_D_N})) show that the output specificity increases with increasing $\phi$. As before, the two requirements of decreasing $\Delta L_{min} (L_s)$ and increasing specificity are incompatible, but increasing $N$ for fixed $m$ relaxes the trade-off.

\section{Consequences of Finite Unbound Receptor Dephosphorylation Rate}\label{sec:non_infinite_unbound_dephosphorylation}

An important assumption of our model is that the dephosphorylation of the internal section of unbound TCR is fast. This leads to all complexes in the signalling cascade decaying to $R^{\scriptsize{\mbox{free}}}$ and $L^{\scriptsize{\mbox{free}}}$ with a rate equal to the inverse binding time (e.g. Fig.~2~(d) of the paper). How specifically our model depends on this assumption needs to be assessed.

\begin{figure}[h!]
\centering
\includegraphics[width=0.8\textwidth]{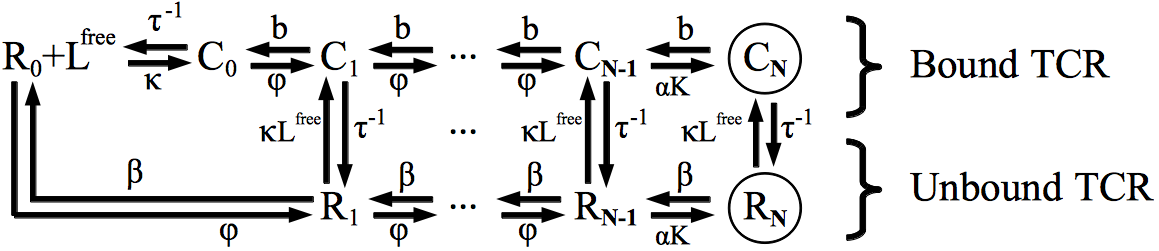}
\caption{\label{fig:finite_unbound_dephospho_model} Parallel adaptive sorting scheme with finite dephosphorylation $\beta$ of unbound TCR. $R_n$ denotes the unbound TCR phosphorylated $n$ times. As in the parallel adaptive sorting module (Fig.~\ref{fig:schematic_generalized_adaptive_sorting}), kinase $K$ is deactivated by $C_m$ and $R_m$ (not shown for clarity). Only one type of ligand is shown.}
\end{figure}

To this end, consider the conservative model where dephosphorylation of the internal section of the unbound TCR occurs distributively (worst case scenario) with finite rate $\beta$. Assuming that the rest of the biochemistry is unaffected, we have the network illustrated in Fig.~\ref{fig:finite_unbound_dephospho_model}.

It is intuitively clear that for $\beta \rightarrow \infty$, the new model reduces to the parallel adaptive sorting scheme described in Sec.~\ref{sec:general_calculation}.   We consider the situation $\beta < \infty$ to understand limitations of our description.

The first thing to notice with $\beta < \infty$ is that even in the absence of pMHC molecules, there will be basal levels of modified TCR (i.e., $R_n > 0$ for all $n$). In the absence of ligands, it is straightforward to show that

\begin{equation}\label{eqn:R_n}
R_n = \left( \frac{1 - \phi \beta^{-1}}{1 - \phi^N \beta^{-N}}\right) \left(\frac{\phi}{\beta}\right)^n R \approx \left(\frac{\phi}{\beta}\right)^n R~~~~~\mbox{for}~~\phi \beta^{-1} \ll 1~~~\mbox{(for}~~n>0).
\end{equation}

In the presence of ligands, the steady-state value of $R_n$ is more complicated although we expect the above expression to be valid at low ligand concentrations. We are interested in determining whether or not the properties of the network is strongly compromised by $\beta < \infty$. One property of particular importance is the fact that the network must be sensitive to low concentrations of foreign ligands. In that regard, note that the equation for the steady-state value of the kinase $K$ becomes in the present case, using equation (\ref{eqn:R_n}) (cf. equation (\ref{eqn:kinase_K}))

\[
K = \frac{K_T}{1 + C_*^{-1} \left( C_m + R_m \right) } \approx \frac{K_T}{1 + C_*^{-1} \left( C_m + \phi^m \beta^{-m} R \right) }.
\]

We see that just as introducing another type of ligand (self ligands, see equation (4) of paper), introducing a finite $\beta$ leads to a shift in the ligand concentration required to reach the adaptive regime. Crudely, taking our low ligand expression for $R_m$, we expect that a finite $\beta$ will lead to a shift in the half-maximum ligand concentration given by  (neglecting a term of order 1 since $\kappa R \tau \gg 1$ for $\tau > 1$~s)
\begin{equation}\label{eqn:delta_L_half_beta}
C_m \gg R_m ~~\Rightarrow~~\Delta L_{1/2}(\beta) \approx \left(\frac{1 + \phi \tau}{\beta \tau}\right)^m \left(1 + \phi \tau\right) R
\end{equation}

\begin{figure}[h!]
\centering
\includegraphics[width=0.85\textwidth]{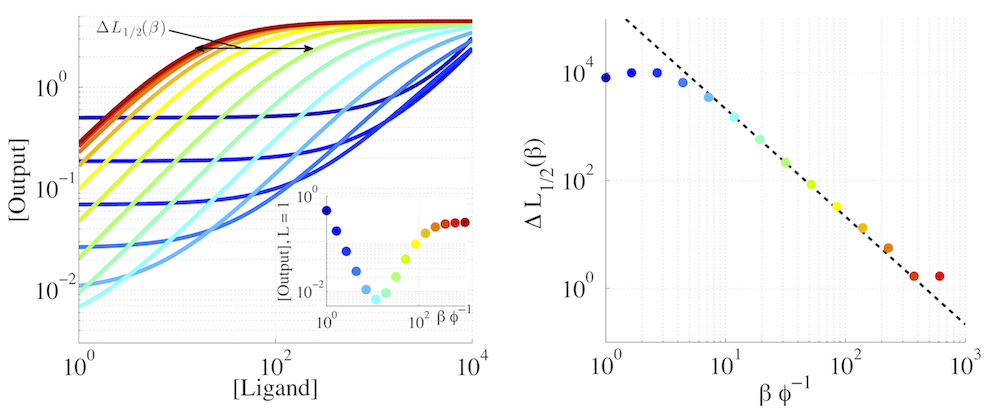}
\caption{\label{fig:finite_unbound_dephospho} Left panel shows output versus ligand relationship for different values of $\beta$ (color coded to right panel). Inset shows the ``basal" output level as a function of $\beta$. Right panel shows the shift in $L_{1/2}$ due to $\beta$ (shown in left panel). Points come from numerical integration and dashed line is equation (\ref{eqn:delta_L_half_beta}). Parameters: $N = 4$, $m = 2$, $\tau = \tau_f = 10$ s, $L_s = 0$, $R = 3\times10^4$, $\kappa = 10^{-4}$, $\phi = \alpha = 0.3$, $b = 0$, $K_T = 1$, $\delta = 1$ and $\epsilon = 0.5$.}
\end{figure}

Figure \ref{fig:finite_unbound_dephospho} illustrate this effect for a given set of parameters. Equation (\ref{eqn:delta_L_half_beta}) captures quite well the effect of a finite $\beta$ on the sensitivity of the model. The larger $\beta \phi^{-1}$, the smaller the decrease in sensitivity. Interestingly, we see from equation (\ref{eqn:delta_L_half_beta}) that increasing $m$ (i.e., moving the position of the activation of the adaptive sorting module downstream) helps to reduce the ``basal" deactivation of the kinase $K$ by unbound TCR. See below for an opposing effect however.

An interesting feature of the network with $\beta < \infty$ is that at low ligand concentration, the output shows a non-monotonic behaviour in $\beta$ (Fig.~\ref{fig:finite_unbound_dephospho}, left panel inset). One way to conceptually think about this is that with finite $\beta$, the unbound receptor behave essentially as another ligand type with binding time $\sim \beta^{-1}$. From equation (\ref{eqn:delta_L_half_beta}), we see that for $\beta > \phi$, the main effect of modified unbound receptor is to desensitize response. This explains the initial decrease in output as $\beta$ decreases. As $\beta$ gets on the same order of magnitude as $\phi$, we see from equation (\ref{eqn:R_n}) that the unbound receptors will actually begin to contribute to the output itself. This explains the increase in output concentration for low $\beta$.

For our model to remain sensitive in the presence of finite unbound dephosphorylation, we must demand that the unbound TCR do not deactivate the response. From equation (\ref{eqn:delta_L_half_beta}), this corresponds to $\Delta L_{1/2}(\beta) \ll 1$, or to $\beta \gg (\phi + \tau^{-1}) \sqrt[m]{R (1 + \phi \tau)}$.

\begin{figure}[h!]
\centering
\includegraphics[width=0.85\textwidth]{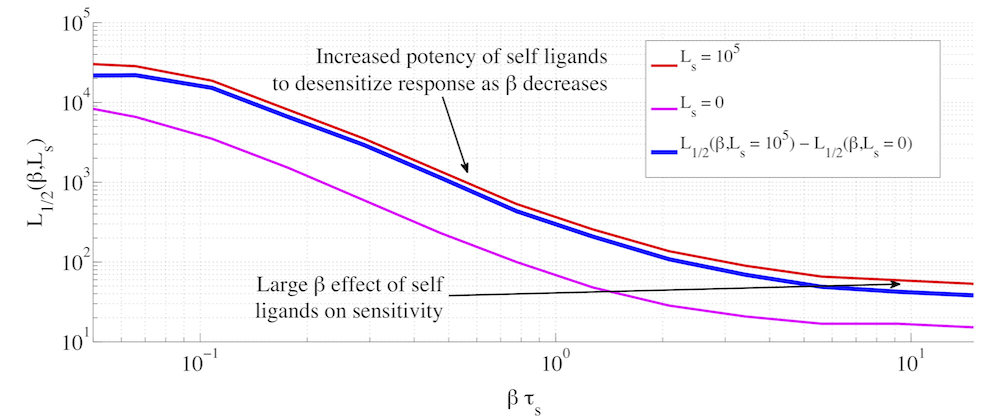}
\caption{\label{fig:delta_L_s_beta} Shown is the loss of sensitivity (quantified through $L_{1/2}$) to foreign output as a function of $\beta$ in the presence (red) and absence of self (magenta, essentially the same data as in Fig.~\ref{fig:finite_unbound_dephospho}). The blue curve shows the behaviour of $L_{1/2}$ arising from self only (subtraction of $L_{1/2}$ with $L_s = 0$ from that with $L_s = 10^5$).  We see that as $\beta$ decreases, the potency of self ligands to desensitize (larger $L_{1/2}$) response increases. Parameters as in Fig.~\ref{fig:finite_unbound_dephospho}, except for $L_s$.}
\end{figure}

It is also important to mention that the potency of self ligands to desensitize response will be enhanced by a finite $\beta$. This is qualitatively easy to understand: a finite $\beta$ leads to more intermediate complexes in the cascade as a result of a slower rate to $R^{\scriptsize{\mbox{free}}}$ and $L^{\scriptsize{\mbox{free}}}$. In particular, a finite $\beta$ will increase $D_m$, the complex arising from self ligands responsible for the deactivation of the kinase. This means that the condition $C_m \gg D_m$ required to reach the asymptotic output concentration will be harder to achieve, implying a lower sensitivity to foreign ligands. This effect arises as $C_m$ is not as strongly affected as $D_m$ by $\beta < \infty$ because $\tau_f \gg \tau_s$. To see this, one can very roughly approximate the effect of a finite $\beta$ as an increase in the binding time of the complexes in the cascade. To return to $R^{\scriptsize{\mbox{free}}}$ and $L^{\scriptsize{\mbox{free}}}$, complex $n$ must not only unbind, but also undergo $n$ dephosphorylations. So, neglecting rebinding possibilities (reasonable for large $\beta$), one has effectively that $\tau \rightarrow \tau + n \beta^{-1}$ for complex $n$, where $\beta^{-1}$ is the mean time needed for one dephosphorylation of an unbound receptor. It follows that the relative effective change in $\tau$ will be more important the smaller $\tau$ is, implying a more pronounced effect of $\beta < \infty$ for complexes arising from self ligands. From this simple estimate, we know that this enhanced desensitization due to self ligands should be small provided $\beta  \gg m \tau_s^{-1}$. Numerical simulations (e.g. Fig.~\ref{fig:delta_L_s_beta}) confirm this. This restriction for the conclusions of our model to be valid is in addition to the previous one arising from equation (\ref{eqn:delta_L_half_beta})).

Note that from Sec.~\ref{sec:non_zero_dephosphorylation}, $b \ll \tau$ was also required for optimal specificity. There thus seems to be a tension between these two requirements: small dephosphorylation rate of unbound TCR but large dephosphorylation rate of bound TCR. However, it must be stressed that this model is fully consistent with the kinetic segregation model \cite{Davis:2006}, which has recently received strong experimental support \cite{James:2012}. In that model, the binding of the pMHC to the TCR approaches the cells' membranes, thereby pushing phosphatases with large extracellular domain away from the region of binding. As the pMHC-TCR complex unbinds, the phosphatase can have access to the TCR again. We thus expect, in a very crude sense, that $\beta$ be large and $b$ be small in a model of early immune signal transduction incorporating kinetic segregation. Complex membrane biophysics would need to be taken into account to do full justice to the kinetic segregation model. This is beyond the scope of the present paper. Our model is in a way a zeroth order approximation consistent with the essence of kinetic segregation.

\section{Robustness of Adaptive Sorting to Stochasticity}\label{sec:noise}

Our discussion thus far was purely in deterministic terms. Given the low concentrations of ligands considered, such perspective is incomplete.

It is crucial to verify that the performance of adaptive sorting is not compromised by stochasticity. In what follows, we perform all our stochastic simulation of the adaptive sorting networks with the Gillespie algorithm \cite{Gillespie:1977}.

\subsection{General Considerations}
In the present work, we are specifically concerned with the steady-state output concentration $O(\tau,L)$ of the signalling pathway, as a function of the ligand concentration and binding time. 

\begin{figure}[h!]
\centering
\includegraphics[width=\textwidth]{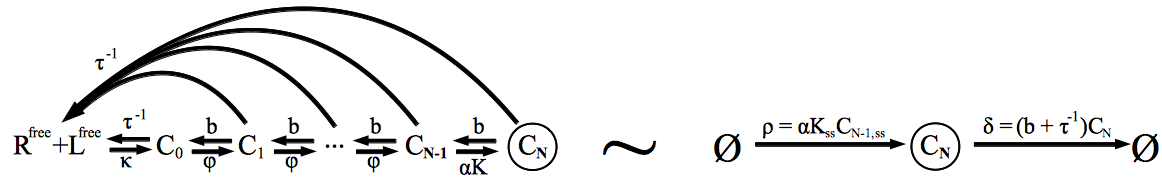}
\caption{\label{fig:stochastic_simplification} We approximate the full network as a simple Poisson birth-death process for the output. This approximation should be valid provided the concentration of other molecules is large compared to the output. The subscript ``ss" denotes steady-state values. $\emptyset$ denotes the fictitious pool of molecules from which $C_N$ is produced and degraded to.}
\end{figure}

In order to make progress in the understanding of the stochastic properties of the displayed solutions, we make the following reasonable assumption for a signalling cascade: The output concentration is lower than the concentration of other species. If this is true, then it is possible to treat relevant effects of fluctuations by focusing specifically on the output. A natural way to do this is to assume that the output follows a Poisson birth-death process, with production ($\rho$) and degradation ($\delta$) rates given by the corresponding deterministic rates. Doing this neglects correlation between the output and other species, but should be valid in the limit that our initial assumption (output concentration smaller than other species' concentration) is valid. This approximation is schematically illustrated in Fig.~\ref{fig:stochastic_simplification} for the specific example of a parallel sorting scheme. 

We can further approximate our Poisson birth-death process by a simple Langevin scheme  \cite{VanKampen:1992}. This approximation is valid at large molecular number, which is not strictly the regime of interest. We need this to make analytical progress. The stochastic differential equation is

\[
\frac{\mbox{d}\tilde{O}}{\mbox{d}t} = -\delta \tilde{O} + \Gamma(t) \sqrt{2\rho + \delta \tilde{O}} \approx -\delta \tilde{O} + \Gamma(t) \sqrt{2\rho} , ~~\mbox{where}~~O = \tilde{O} + \rho \delta^{-1}.
\]

$\Gamma(t)$ is Gaussian white noise with unit variance. The initial condition is that $\tilde{O}(t = 0) = -\rho \delta^{-1}$. By construction, we have that $\rho \delta^{-1} =  O^{\scriptsize{\mbox{det}}}(L)$, the deterministic steady-state output concentration. It is important to stress that in our scheme this is independent of the specific topology or parameters of the underlying network. The second approximate equality above comes from neglecting $\tilde{O}$ in the square root. Since in the steady-state, $\langle \tilde{O} \rangle = 0$, this amounts to doing a first order approximation. $\langle \cdot \rangle$ denotes an ensemble average.

In the end, we see that the full network can be approximated by a Ornstein-Uhlenbeck process, whose stochastic properties are well known. In particular, note that \cite{Gillespie:1996}
\begin{equation}\label{eqn:mean_C_N_tilde}
\langle \tilde{O}(t) \rangle = -  O^{\scriptsize{\mbox{det}}}(L)  e^{-\delta t},
\end{equation}

\begin{equation}\label{eqn:auto_correlation_C_N_tilde}
\langle \tilde{O}(t)~\tilde{O}(t') \rangle = \langle \tilde{O}(t)\rangle\langle\tilde{O}(t') \rangle + O^{\scriptsize{\mbox{det}}}(L) e^{-\delta (t-t')} \left( 1 - e^{-2\delta t'} \right)~~(t \geq t').
\end{equation}

Since we are considering low number of output molecules, the output's time course will be highly fluctuating on short time scales. As a result, it is not possible to make decisions concerning the nature of the bound ligands (critical or foreign) given observations of the output alone.

Motivated by the biological system, we introduce species $A$ downstream of the output responsible for time averaging the output \cite{Francois:2013}. In our model, the thresholding mechanism acts on $A$, not on $O$. It is reasonable to assume that the discrimination and amplification steps are realised by different modules in the actual biological system. We take
\[
\dot{A} = \Lambda O - T^{-1} A,
\]
where $T$ is effectively the averaging time. We suppose in what follows that the kinetics of $A$ is slow compared to that of the output ($\delta T \gg 1$). $\Lambda$ is a kinetic first order constant that sets the scale of concentration $A$. Many molecular species could play the role of $A$ in signal transduction of the immune system and most of them have typical concentrations above $\sim 10^4$ \cite{AltanBonnet:2005}. It is then reasonable to assume that $\Lambda$ is large enough that stochasticity of $A$ will be entirely dominated by the stochasticity of the output $O$. We therefore integrate for $A$ deterministically. With $A(0,L) = 0$, we can formally write the solution of $A(t, L)$ (note that $A$ depends on the ligand concentration) as function of $O(t,L)$,
\begin{equation}\label{eqn:solution_A}
A(t,L) = \Lambda \int_0^t e^{-(t-t')/T} O(t',L) dt'.
\end{equation}

With our approximations, equations (\ref{eqn:mean_C_N_tilde}), (\ref{eqn:auto_correlation_C_N_tilde}) and (\ref{eqn:solution_A}) together with our definition of $\tilde{O}$ allow us to compute the stochastic properties of $A(t, L)$. One finds (recalling that $\rho \delta^{-1} = O^{\scriptsize{\mbox{det}}}(\tau,L)$, the deterministic steady-state output concentration) 
\begin{equation}\label{eqn:mean_A}
\langle A(t,L) \rangle = O^{\scriptsize{\mbox{det}}}(L)\Lambda T \left\{1 - e^{-t/T} + \tilde{\tau} e^{-t/T} \left(1 - e^{-t/\tilde{\tau}} \right) \right\}~~~\mbox{where}~\tilde{\tau} = T  \left(\delta T - 1\right)^{-1},
\end{equation}

\begin{equation}\label{eqn:std_A}
\sigma_{A(t, L)}^2  = \langle \left(A(t, L)\right) ^2\rangle - \langle A(t, L)\rangle^2 \approx O^{\scriptsize{\mbox{det}}}(L) \left(\frac{\Lambda^2 T^2}{T \delta + 1}\right) \left( 1 - e^{-2t/T}\right).
\end{equation}

We neglected transient terms smaller by a factor of $\delta^{-1} T^{-1}$ in our expression for $\sigma_{A(t,L)}^2$. Given the level of approximation we are after, the exact form of the variance of $A(t,L)$ is not more useful than the above compact expression. From these results, we can see for what value of $T$ discrimination between critical and foreign ligands become possible with parallel adaptive sorting.

To do so, we must first determine a plausible threshold value. Optimally, the threshold should be positioned at a value as low as possible (to be sensitive to as little foreign ligands as possible), but not sufficiently low as to trigger response with critical ligands (no false positive responses). Since the output increases monotonically with ligand concentration in an adaptive sorting scheme, we can consider the large $L_c$ limit for $O^{\scriptsize{\mbox{det}}}(L_c)$, denoted by $O_c^{\scriptsize{\mbox{det}}}$. We can set our threshold $\theta$ to be
\begin{equation}\label{eqn:threshold_A}
\theta(\psi) = \langle A(\tau_c) \rangle + \psi \sigma_{A_c} = \Lambda T \left( O_c^{\scriptsize{\mbox{det}}} + \psi \sqrt{\frac{O_c^{\scriptsize{\mbox{det}}}}{T\delta_c + 1}}\right) ,
\end{equation}
where $\psi$ quantifies how conservative the system needs to be about auto-immunity. $A_c$ above denotes the large time and ligand concentration limit of $A$ for critical ligands. Larger $\psi$ implies less frequent false positives. The underlying biochemical network will provide a suitable discrimination mechanism even in the presence of noise provided one can still find
\[
Z(\psi,L_f) = \frac{\langle A(L_f)\rangle - \theta(\psi)}{\sigma_{A(L_f)}} 
\]
on the order of one or larger for even very small foreign ligand concentrations. $Z(L_f)$ tells us how far from the threshold (measured in units of typical fluctuations of $A$) the output arising from foreign ligands is at concentration of $L_f$. Indeed, from equations (\ref{eqn:mean_A}), (\ref{eqn:std_A}) and the definition of the threshold $\theta$, equation~(\ref{eqn:threshold_A}), one gets
\begin{equation}\label{eqn:z_factor}
Z(\psi,L_f) \approx  \sqrt{O^{\scriptsize{\mbox{det}}}(L_f) \left(1 + T\delta_f \right)} \left\{ 1 - \frac{O_c^{\scriptsize{\mbox{det}}}}{O^{\scriptsize{\mbox{det}}}(L_f)}\right\} - \psi \sqrt{\frac{\delta_f O_c^{\scriptsize{\mbox{det}}}}{\delta_c O^{\scriptsize{\mbox{det}}}(L_f)}}
\end{equation}
It is approximate only as it neglects terms of order $\delta T^{-1}$. We can evaluate this quantity knowing the model for the signalling cascade. It is worthwhile to substitute numbers. In our model, $\delta$ is essentially the unbinding rate $\tau^{-1}$. Taking $O_c^{\scriptsize{\mbox{det}}} = 0.5$ and $ O_f^{\scriptsize{\mbox{det}}}(L_f = 5) = 1$ are conservative estimates within our models. With $\tau_f = 10$ s and $\tau_c = 3$ s, we have $Z \approx 0.5 \sqrt{1 + 0.1 T} - 0.4\psi$. For $T = 60$ s and $\psi = 3$, we already have $Z \approx 0.1$, meaning that we expect more than half of the cells to respond in a time-scale of one minute when exposed to on the order of 5 foreign ligands.

The important point to realize in equation~(\ref{eqn:z_factor}) is that as $T$ increases, the relative importance of $\psi$ (encoding the conservatism of the system against auto-immune reaction) decreases. This simply makes explicit the fact that relative fluctuations decrease as $T^{-1/2}$. One cannot take $T$ to be arbitrarily large, as this would lead to response time too long to be useful in a biological system. However, numerical simulations (next sections) show that $T \sim 60$ s is an appropriate compromise between speed and time averaging of fluctuations. Interestingly, this is precisely the time scale at which cells of the immune system respond in the presence of foreign ligands~\cite{AltanBonnet:2005}.

One might object that for low concentrations, the output will fluctuate above and below the threshold. This could in principle incapacitate the downstream thresholding mechanism. These fluctuations occur on time scales that are slow ($T$, the relaxation time of $A$). There is good empirical evidence that once the molecular equivalent of $A$ is exceeds a threshold in T-cells, further molecular reactions actually deactivate the incoherent feed-forward loop \cite{AltanBonnet:2005} (i.e., reactivating the kinase $K$ in our model). This in turn leads to a sharp increase in the output, making fluctuations irrelevant. It is reasonable that the kinetics of this additional reaction are faster than that of $A$. As a result, this means that response would be triggered provided the output stayed above threshold for a sufficiently long time, dictated by the kinetics of the mechanism downstream of $A$. An approximation is then to take response to occur as soon as the output exceeds threshold. This is the strategy used to generate Fig.~4~(d) of the paper. Observe that such strategy makes the system potentially vulnerable to critical ligands. In spite of this, very few cells ($\sim 1$\% in our time window) respond to critical ligands, even at large concentrations (see Fig.~4~(d) of the paper). In the examples below, we instead display the fraction of cells above threshold at a given time (i.e. not assuming anything about the time-scale of the thresholding mechanism). The qualitative behaviour of the fraction of cells above threshold is very similar to the fraction of cells responding just as threshold is crossed. Note that the downstream mechanism is not explicitly taken into account in our model for reasons of simplicity. 

Many approximations were needed to arrive at our conclusions. In the following sections, we verify numerically for the parallel adaptive module that the conclusions hold. Note that for the module activated by the phosphatase (network from Fig.~4(b) in the paper), an analysis of noise was performed in \cite{Francois:2013}. It was seen that an adaptive sorting like module (but with a different network topology) could have good discriminatory abilities even at low molecular numbers. These results indicate that the conclusions drawn from our deterministic results do indeed remain valid even in the presence of fluctuations.

\subsection{Specific Example - Parallel Adaptive Sorting}

\begin{figure}[h!]
\centering
\includegraphics[width=0.9\textwidth]{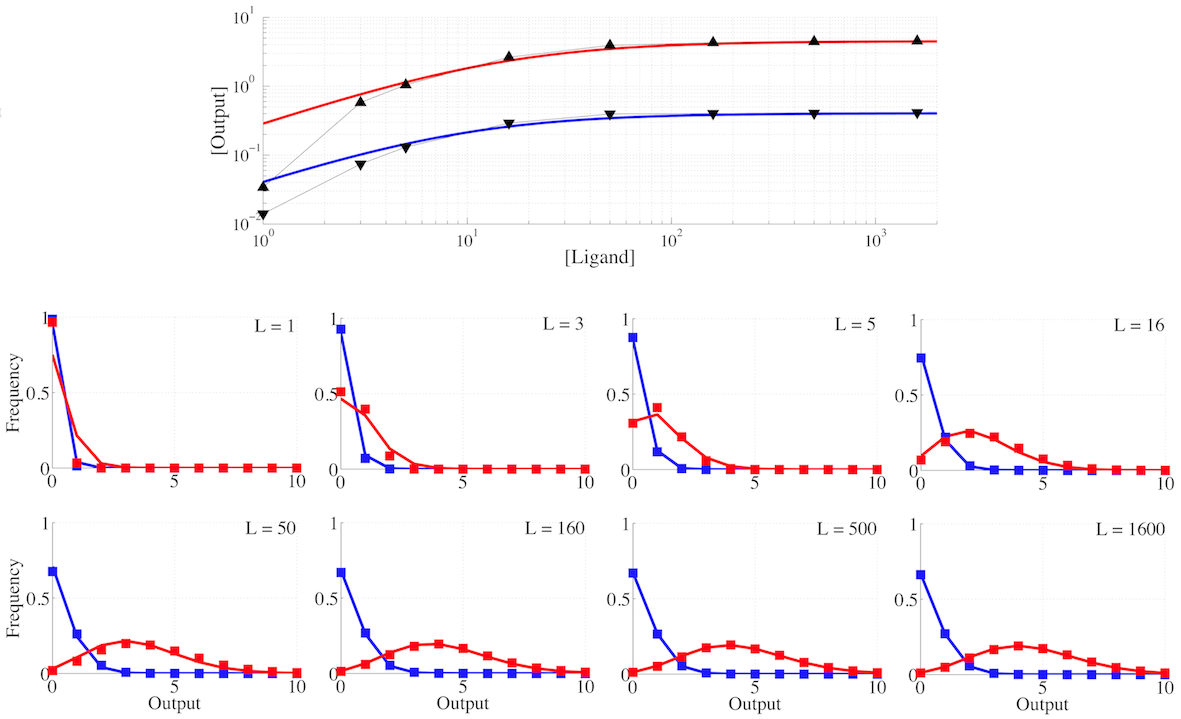}
\caption{\label{fig:stochastic_numerical} Top graph compares deterministic (full lines) and stochastic steady-state output concentrations ($\blacktriangle$ for $\tau_f$, $\blacktriangledown$ for $\tau_c$).  Lower rows display the output number distribution distributions (squares) for the ligand concentrations of the top graph. Poisson distributions (full lines) with the analytical deterministic result for the mean of $C_N$ (no fit). Observe how the Poisson distribution with the deterministic mean becomes a valid approximation as $L$ increases. Even at $L = 5$, the Poisson distribution captures well the output distribution. Parameters: $N = 4$, $m = 2$, $L_s = 0$, $R = 3\times10^4$, $\kappa = 10^{-4}$, $\phi = 0.3$, $\alpha = 0.0003$, $b = 0$, $K_T = 1000$, $\delta = 1$ and $\epsilon = 0.5$.}
\end{figure}

We here consider the idealized network performing parallel adaptive sorting (network from Fig. 4 (a) from the paper), first in the absence of self ligands ($L_s = 0$). The first approximation of importance is that the output concentration is governed by a Poisson process with production and degradation rates equal to the deterministic ones. The top of Figure~\ref{fig:stochastic_numerical} compares the deterministic and stochastic mean concentration in the steady-state (we take a unit volume). 

As expected, the stochastic mean output number differs significantly from the deterministic output at very low ligand concentrations. We see however that there is good agreement even for $L \sim 3$. The bottom of Figure~\ref{fig:stochastic_numerical} compares the output molecule number distributions to Poisson distribution with rates from the deterministic network (note that the lines are not fit).

\begin{figure}[h!]
\centering
\includegraphics[width=0.8\textwidth]{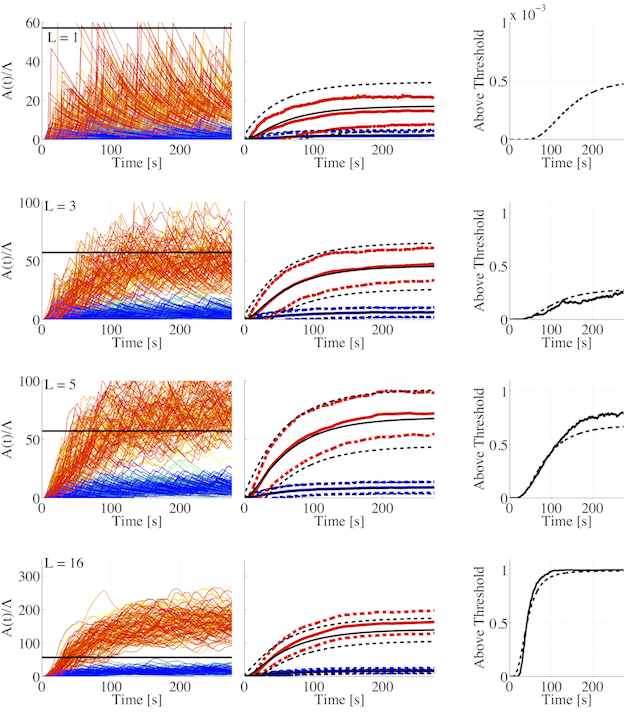}
\caption{\label{fig:final_summary_stochastic} Each row represent a different ligand concentration (same concentrations as in Fig.~\ref{fig:stochastic_numerical}). The left panels present representative time evolutions of $A(t)$. Lines of warm color are with $\tau_f$ and lines of cool colors with $\tau_c$. Horizontal black line is the threshold chosen from equation~(\ref{eqn:threshold_A}) with $\lambda = 4$. It is the same for all rows.  Averaging time $T$ is taken to be 60 s. Middle panels show the mean and standard deviations of the trajectories from numerical results (blue $\tau_c$, red $\tau_f$) and equations (\ref{eqn:mean_A}) and (\ref{eqn:std_A}) (black). Ensemble average over 500 trajectories. Right panels show the fraction of trajectories above threshold at a given time (showing those for $\tau_f$, not more than few trajectories with $\tau_c$ exceeded sporadically threshold). Note that the above is a different quantity than that presented in Fig.~4~(d) of the paper. See discussion. Full lines are numerical and dashed lines are from equations (\ref{eqn:mean_A}) and (\ref{eqn:std_A}). Too few trajectories exceeded threshold to have proper statistics for $L=1$ (note the scale). Observe the relatively short timescale over which the fraction increases. Kinetic parameters are as in Fig.~\ref{fig:stochastic_numerical}. }
\end{figure}

\begin{figure}[h!]
\centering
\includegraphics[width=0.8\textwidth]{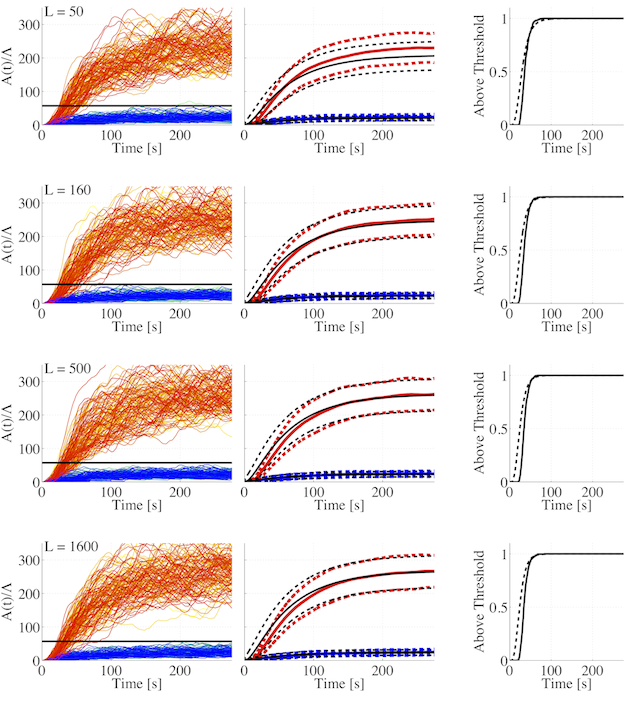}
\caption{\label{fig:final_summary_stochastic_II} Continuation of Fig.~\ref{fig:final_summary_stochastic} with remaining concentrations. Observe that $L = 160,~500$ and $1600$ are practically indistinguishable, as we expect from the main idea of adaptive sorting that output should not depend on concentration at large ligand concentrations. }
\end{figure}

For the same parameters, we can also verify that the statistical properties of $A$ as derived compare well to the full stochastic result. Sample trajectories for $A$ as well as $\langle A(t,L) \rangle \pm \sigma_{A(t,L)}$ as given by equations~(\ref{eqn:mean_A}) and (\ref{eqn:std_A}) with deterministic output given by equation~(\ref{eqn:final_result}) are shown in Fig.~\ref{fig:final_summary_stochastic} and \ref{fig:final_summary_stochastic_II}.

Note that at moderately high concentrations ($L = 160$, second row in Fig.~\ref{fig:final_summary_stochastic_II}), the analytic expression is very close to the actual numerical result. This is expected as the correlations between the output and other species should decrease in importance as the concentration of ligands increases. Agreement of steady-state is also quite good at other concentrations in spite of our numerous approximations (there are discrepancies at intermediate ligand concentrations). 

\begin{figure}[h!]
\centering
\includegraphics[width=0.7\textwidth]{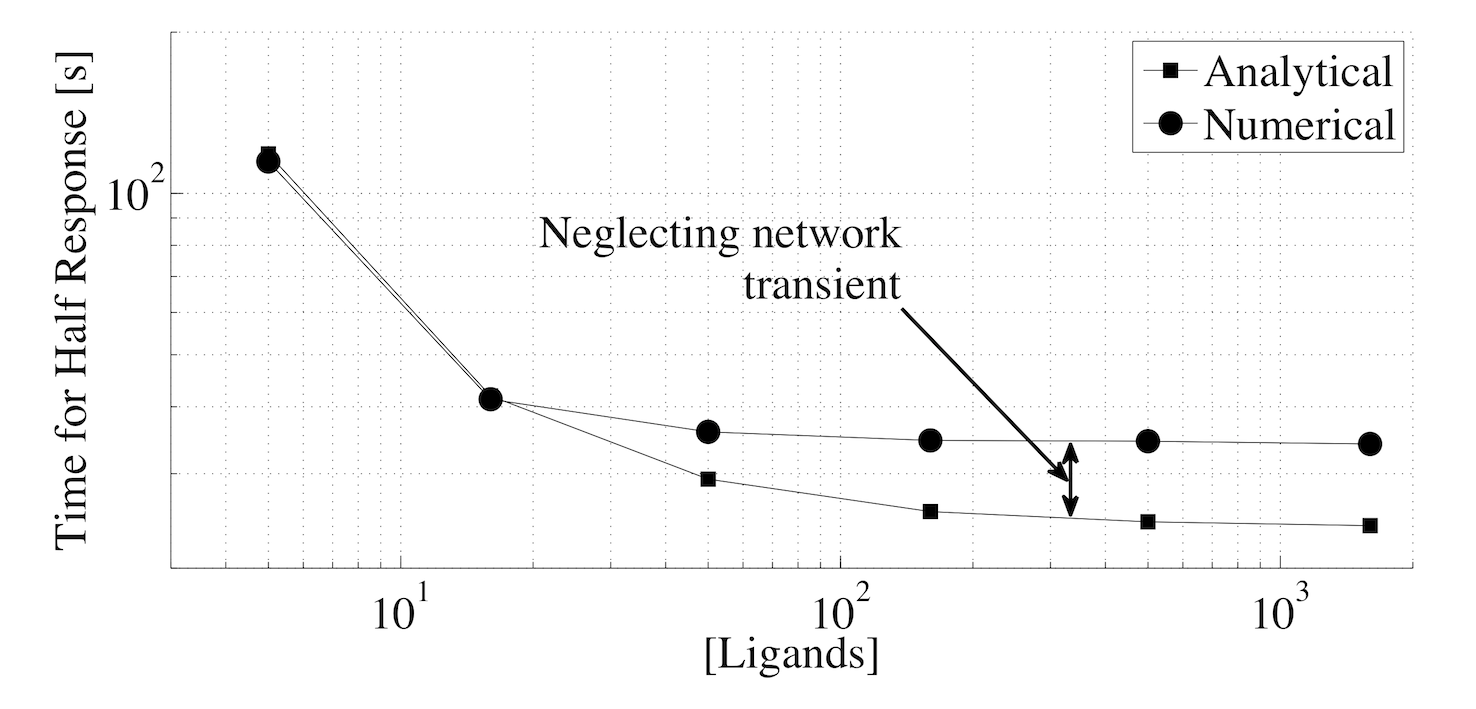}
\caption{\label{fig:half_times} Time required for half of trajectories to exceed threshold (for foreign ligands) as a function of ligand concentration. Note the systematic discrepancy of our analytic approximation at large ligand concentrations. The analytic result however recovers the qualitative behaviour of the full system. Same parameters as in Fig.~\ref{fig:final_summary_stochastic} and \ref{fig:final_summary_stochastic_II}. }
\end{figure}

Systematic discrepancies in the transient behaviour is seen for all concentrations: the analytic solution increases too fast at early times. This can be understood simply. In our approximation of the full network, (shown in Fig.~\ref{fig:stochastic_simplification}), we neglected the transient of all the species but the output. Specifically, we take the rates governing the Poisson process for the output to be constant. Clearly, there is a relaxation time (on the order of $\tau$) for these rates to actually attain their steady-state value. Therefore, the output steady-state will be attained slower in the full network than in our approximation. One way to quantify the discrepancy is to plot the time required for half the trajectories to be above the threshold. This is shown in Fig.~\ref{fig:half_times} for the concentrations and parameter values of Fig.~\ref{fig:final_summary_stochastic} and \ref{fig:final_summary_stochastic_II}.

Notice how the time for response decreases until a plateau. It is expected in our model that extremely high concentrations of foreign ligands do not trigger the response faster than moderate concentrations (by construction of adaptive sorting). This is actually seen in experiments (\cite{AltanBonnet:2005}, see Fig.~3 (c)). In our model, what should decrease the response time is a more ``potent" foreign ligand, with higher binding time.
 
We ran the same simulations in the presence of self ligands: $L_s = 10^4$ with $\tau_s = 0.11$ s. (In our model, this is as ``potent" as $L_s = 10^5$ with $\tau_s = 0.05$ s. We need this to speed up the Gillespie simulations). We see that just as in the $L_s = 0$ case, the Poisson approximation is excellent even at very low concentration. Our analytic result also capture quite well the exact stochastic behaviour. We show below $Z(L_f)$ (equation (\ref{eqn:z_factor})) in the presence and absence of self. Note again that while not perfect, our analytical results capture very well the full stochastic behaviour. 

\clearpage

\vspace{5mm}

\begin{figure}[h!]
\centering
\includegraphics[width=0.8\textwidth]{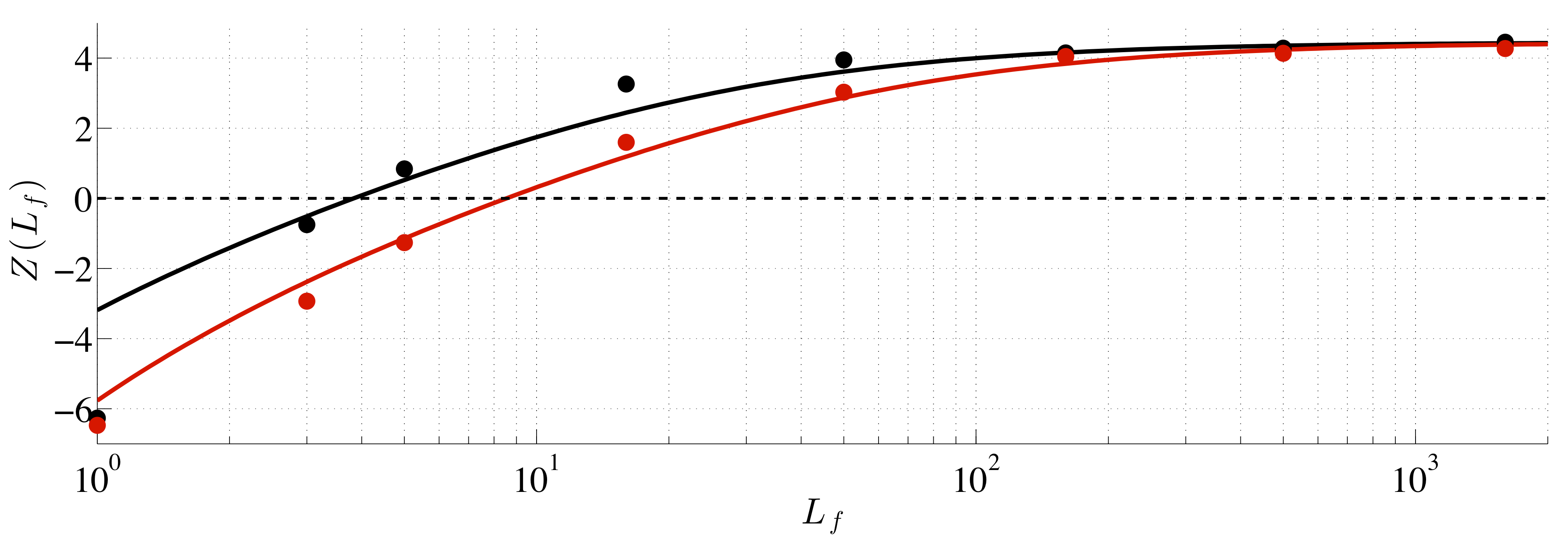}
\caption{\label{fig:z_factor} $Z(\psi,L_f)$ from equation (\ref{eqn:z_factor}). Black line for $L_s = 0$ and red line for $L_s = 10^4$ ($\tau_s = 0.11$ s). Dots are from full stochastic integrations. Parameters from Fig.~\ref{fig:stochastic_numerical}, \ref{fig:final_summary_stochastic} and \ref{fig:final_summary_stochastic_II} ($\psi = 4$, $T = 60$ s). Note that already at $L_f = 4$, there should be $\sim$ 0.5 of response in the absence of self. In the presence of self, this goes to $L_f = 9$, which is still a very small number.}
\end{figure}

\begin{figure}[h!]
\centering
\includegraphics[width=\textwidth]{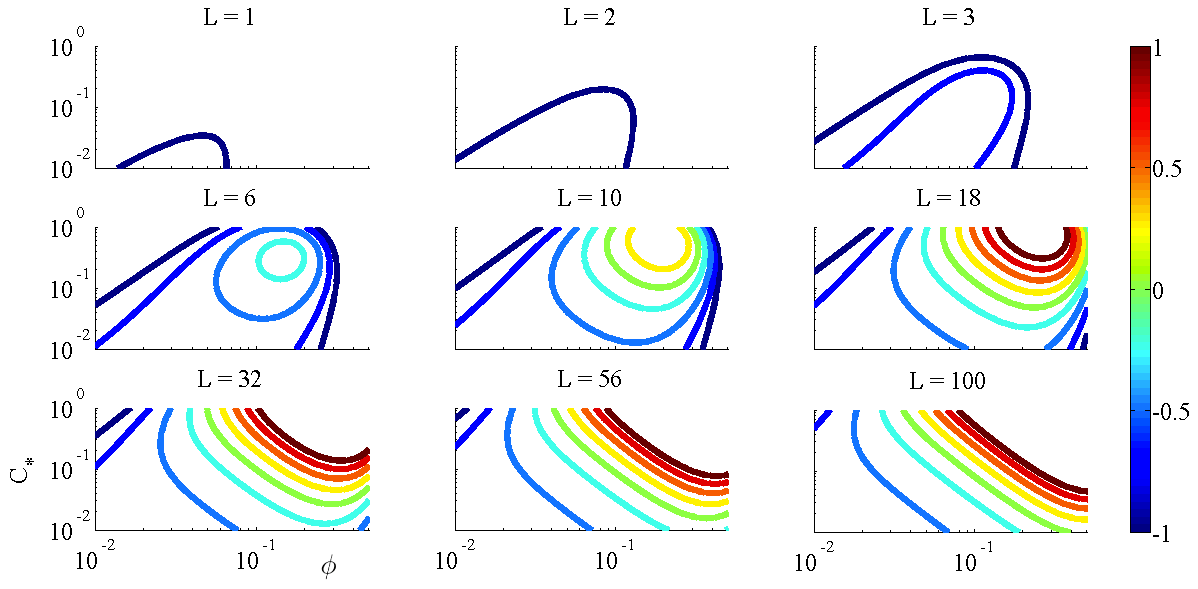}
\caption{\label{fig:z_factor_many_L} $Z(\psi,L_f)$ from equation (\ref{eqn:z_factor}) for parallel adaptive sorting module with $N = 4$, $m = 2$, $\kappa R = 3$, $L_s = 10^5$, $\tau_s = 0.05$ s, $\alpha K_T = \phi$, $\tau_f = 10$ s, $\tau_c = 3$ s. Varied parameters are $\phi$ (horizontal axis) and $C_*$ (vertical axis). Averaging time 60 s. $\psi = 4$. Different graphs correspond to different foreign ligand concentrations. The green contour corresponds to 50\% response. Dark blue curve to $\sim 16$\% response and dark red to $\sim 86$\% response.}
\end{figure}

From our analytical expression for $Z$, we can quickly assess the robustness to parameter choice for the parallel adaptive sorting. Figure~\ref{fig:z_factor_many_L} shows contours of $Z$ for different parameters and different values of foreign ligands. We see that there is a wide range of parameters where substantial response occurs (here with $T = 60$ s).

\clearpage

\section{Analysis of Transient Behaviour}\label{sec:transient}

We mostly focused with the behaviour of the steady-states in our analysis. This requires justification. It important to verify whether the transient has properties that could potentially compromise our conclusions. We consider the simple adaptive triage module.
We are now interested in the full time evolution of (we assume only one type of ligand) equation~(\ref{eqn:simple_solution}). Exact analysis is not tractable. We obtain approximate expressions. First however, $\Sigma \equiv C_0 + C_1$ can be integrated exactly as
\[
\Sigma(t) = \frac{\kappa R \tau L}{1 + \kappa R \tau}\left(1 - e^{-t/\tau_\Sigma}\right),~~~~\mbox{where }~\tau_\Sigma = \frac{\tau}{1 + \kappa R \tau} \sim 1 \mbox{s}.
\]

 We now consider the equation for the kinase $K$. Note that in the present model, $C_{1,ss}\ll~C_{0,ss}$ (ss, steady-state) at moderately large concentration of ligand. It is then a fair approximation (to be verified at the end) to take $C_0 \sim \Sigma$ in the equation for $K$:

\[
\dot{K}(t) \approx -\left(\delta \Sigma(t) +\epsilon\right) K(t) + \epsilon K_T 
\]
It is possible but not very useful to formally write down a solution for $K(t)$. What is important to realise is that since $\Sigma$ increases monotonically, the rate of decay of $K(t)$ increases monotonically. For the purpose of determining the behaviour of $C_1$ (the output), we only wish to determine when $K(t)$ is close to $K_T$ and when it is close to its (much lower) steady-state value. It is not hard to show then that the following qualitatively captures the behaviour of the kinase:
\[
K(t) \sim K_T e^{-\delta \eta t+\delta \tau_\Sigma \Sigma(t)} + \frac{K_T \epsilon}{\epsilon + \delta C_{0,ss}},
\]

where $\eta \equiv  \lim_{t \to \infty} \Sigma(t) = \kappa R \tau L (1 + \kappa R \tau)^{-1}$. This approximation is valid provided $\delta \eta \tau_\Sigma \gg 1$.

We can turn to the behaviour of the output, which is our primary concern. Assuming as before that $C_0 \sim \Sigma$ simplifies the equation to
\[
\dot{C}_1(t) \approx \alpha K(t) \Sigma(t) - \tau_{C_1}^{-1} C_1(t),~~~~~~~\mbox{where }~\tau_{C_1} = \tau (1 + b \tau)^{-1}.
\]

The first term on the right hand side represents a time varying rate of production for $C_1$. The second term represents a degradation term with a constant half-life. The important point is that since $K(t)$ decreases and $\Sigma(t)$ increases, the production rate will be peaked at some intermediate time value. Integrating the equation for $C_1$ leads to

\[
C_1(t) =  \left(\frac{\alpha K_T \epsilon \eta}{\epsilon + \delta C_{0,ss}} \right) t + \alpha K_T  \int_0^t dt' \left( e^{-\delta \eta t'+\delta \tau_\Sigma \Sigma(t')} \Sigma(t') - \frac{\epsilon \eta e^{-t'/\tau_\Sigma}}{\epsilon + \delta C_{0,ss}}  \right) - \frac{1}{\tau_{C_1}} \int_0^t C_1(t).
\]

It must be stressed that the integrand of the second term decays for times larger than $\tau_\Sigma$. If we are interested in the behaviour of $C_1$ at large times, then this second term is just a constant which can be evaluated by the method of steepest descent. The remaining equation is then straightforward to solve (using that $\eta \sim C_{0,ss}$, and that $C_{0,ss} \gg \frac{\epsilon}{\delta}$). One arrives at an approximate result for $C_1(t)$,

\begin{equation}\label{eqn:approximate_C1}
C_1(t) \approx \frac{\alpha K_T}{\delta} \left( 1 - \epsilon\tau_\Sigma \right) e^{-t/\tau_{C_1}} + \frac{\alpha K_T \epsilon \tau_{C_1}}{\delta}  \left(1 - e^{-t/\tau_{C_1}} \right)~~~~~~~\mbox{for }~ t \gg \tau_\Sigma.
\end{equation}

\begin{figure}[h!]
\begin{center}
\includegraphics[width=\textwidth]{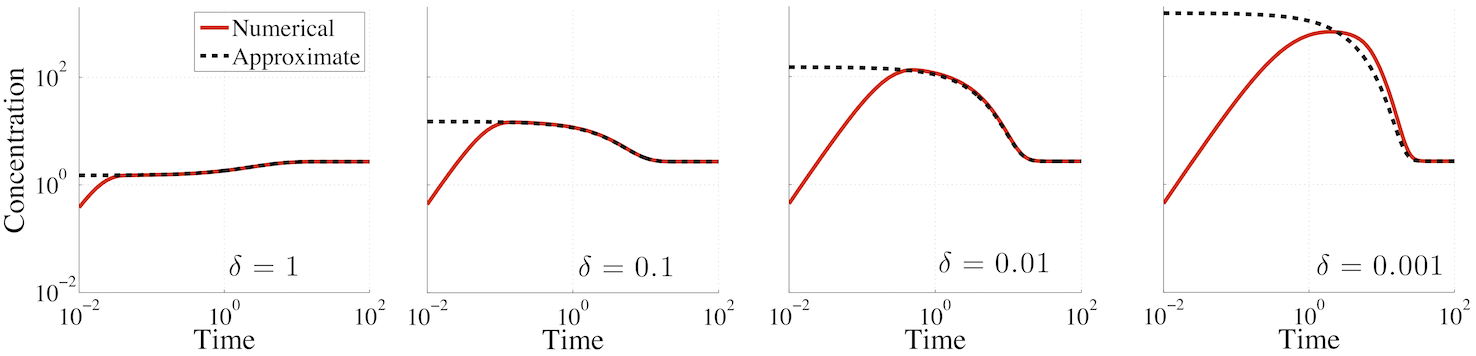}
\caption{\label{fig:transient_examples}Shown in red are numerical integration results for $C_1(t)$ for different values of $\delta$ (with fixed ratio $\frac{\epsilon}{\delta}$ to have the same steady state). The black dashed lines are equations~(\ref{eqn:approximate_C1}). Here, $\frac{\alpha K_T}{\eta} \sim 0.003$ and $\frac{1}{\tau_{\Sigma} \eta} \sim 0.004$. Even out of the range of validity (e.g. $\delta = 0.001$), the approximation still captures qualitatively behaviour of the transient.}
\end{center}
\end{figure}

The first term captures the decay of the initial peak concentration due to the time-varying production rate. The second term represents the relaxation to the equilibrium steady-state. It is important to restate the assumptions used to derive the above expression: (1) $\delta \eta \tau_\Sigma \gg 1$ to obtain our approximate expression for $K(t)$ and (2) $C_1 \ll C_0$ at all times to use $\Sigma(t)$ in place of $C_0(t)$ in our computations. The latter assumption is self-consistent if the peak value of $C_1$ is much smaller than the asymptotic value of $C_0$: $\alpha K_T \ll \eta\delta$. Within these restrictions, numerical integration agrees with the derived result. Examples are shown in Fig.~\ref{fig:transient_examples}.

The final expression for $C_1$ allows us to check the self-consistency of our approximations. We see that the larger $\delta$, then the smaller peak value of $C_1$. This means that our two assumptions are compatible. It is clear however that for very small ligand concentrations, $\eta$ will be small and our approximations break down.

One must keep in mind though that the magnitude of $C_1$ is bounded above by $L$. Hence, the maximum value of $C_1$ cannot be significant at low $L$. This means that even though our approximations do not hold in this regime, only the steady-state behaviour is relevant.

The whole point of this analysis is to verify that the transient of the output does not have a significant behaviour. From our calculations, we see that it can indeed be significant depending on the choice of parameters. In particular, for small $\delta$ (scale set by $\alpha K_T \eta^{-1}$ and $\eta^{-1} \tau_{\Sigma}^{-1}$), $C_1$ exhibits a peak that can have a large magnitude. This implies that for small $\delta$, not only the steady-state is of importance. In particular, if the immune response is triggered as soon as the output reaches a prescribed value (instead of time averaging), then the transient is more important than the steady-state.

More importantly, observe that the peak value of $C_1$ does not strongly depend on $\tau$. This means that the network looses all its fine $\tau$ discriminatory properties if $\delta$ is small. It follows that for our model to work, the kinetics of the kinase must be faster than the kinetics of the phosphorylation of the complex cascade. This fast kinetics ensures a fast decay of $K$, which in turn implies that no significant amount of $C_1$ can accumulate at short times. In such situation, the important behaviour is in the steady state.

\section{Examples of Evolved Networks}\label{sec:evolved_networks}

This section presents networks obtained from our simulations. The raw networks are often overly complicated. For clarity, the essential reactions are extracted and shown in simplified networks (reactions that can be removed with no effect on the output behaviour are not shown).

\subsection{Independent Discrimination - Another Example}
\begin{figure}[h!]
\begin{center}
\includegraphics[width=0.3\textwidth]{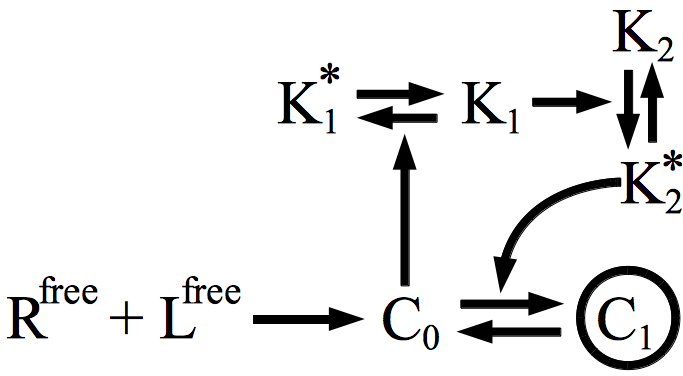}
\caption{\label{fig:det_ex_2_simple_net} Other example of network found. $C_i$'s are understood to decay to $R$ and $L$ with rate $\tau^{-1}$.}
\end{center}
\end{figure}

\begin{figure}[h!]
\begin{center}
\includegraphics[width=0.65\textwidth]{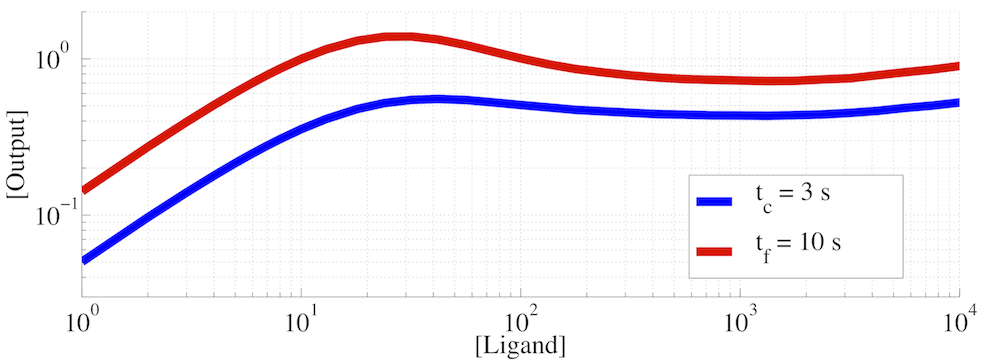}
\caption{\label{fig:det_ex_2_ol}Output versus ligand relationship (steady-state) for network of Fig.~\ref{fig:det_ex_2_simple_net}. Note the non-monotonicity of the output with ligand concentration.}
\end{center}
\end{figure}

We now display another example of the results obtained via simulation, with only one type of ligand present. The simplified network is shown in Fig.~\ref{fig:det_ex_2_simple_net}. The output versus ligand relationship is shown in Fig.~\ref{fig:det_ex_2_ol}.

As for adaptive sorting, $C_0$ is the enzyme phosphorylating some kinase ($K_1$). In contrast to adaptive sorting, $K_1$ does not directly phosphorylate $C_0$. Instead, $K_1$ is required for the formation of $K_2^*$, which is the kinase responsible for the formation of $C_1$ (the output). Hence, more $C_0$ implies less $K_1$ which in turn implies less $K_2^*$, which finally leads to less $C_1$. This is again the same negative feed-forward loop, except with a less direct, "buffered" step. Interestingly, as a consequence, the adaptation is not perfect and the response (steady-state output concentration) becomes non-monotonic in ligand concentration as shown in Fig.~\ref{fig:det_ex_2_ol}. The adaptive sorting module is nevertheless clearly present.

\subsection{Discrimination in Presence of Self Ligands}
Here we show results of evolutionary simulations when discrimination must be achieved in the presence of a large quantity of self ligands (spurious, non-specific ligands).

\subsubsection{Example 1 - Phosphatase Activation}\label{sec:phosphatase_activation}

An example of working obtained network is schematized in Fig.~\ref{fig:par_ex_1_simple_net} (same as Fig.~3~(e) in the paper). The full network and parameters can be found in Appendix A.

\begin{figure}[h!]
\begin{center}
\includegraphics[width=0.7\textwidth]{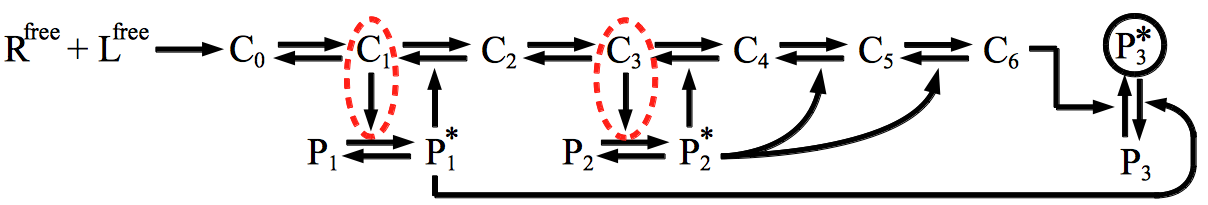}
\caption{\label{fig:par_ex_1_simple_net} Obtained network performing parallel sorting. Complexes $C_i$ are understood to decay to $R$ and $L$ with rate $\tau^{-1}$.}
\end{center}
\end{figure}

Observe first that it is not $C_0$ which activates the crucial ``adaptive reactions" but rather $C_1$ and $C_3$ (dashed circles). The activation of the adaptive module is rewired downstream. Both these reaction serve to activate a phosphatase instead of de-activating a kinase (it is still a negative feed-forward). In fact, this evolved solution is very close to the realistic  model for early immune response presented in \cite{Francois:2013}. Note that the output is not a member of the cascade of complexes, but is rather activated by the last element of the cascade (here $C_6$). This is basically equivalent to having the output as $C_6$ directly. The steady-state output versus ligand relationship is presented as Fig.~3 (d) in the paper.

\begin{figure}[h!]
\begin{center}
\includegraphics[width=0.6\textwidth]{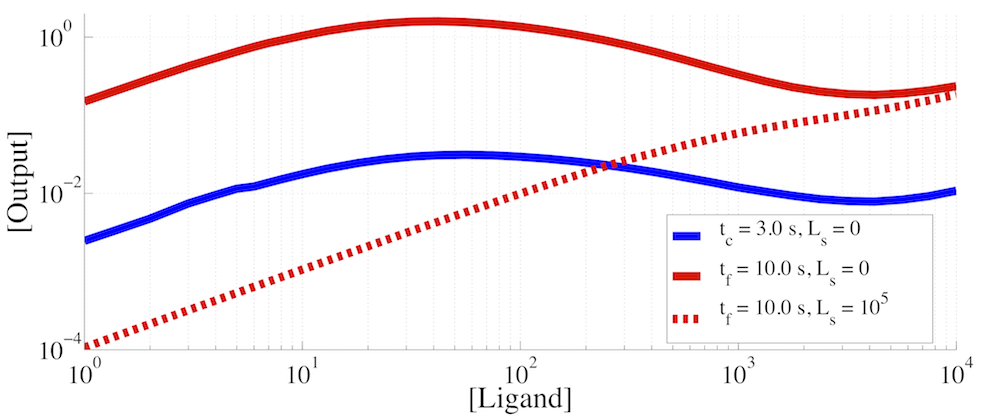}
\caption{\label{fig:par_ex_1_ol_m0} Output versus ligand relationship (steady-state) for network of Fig.~\ref{fig:par_ex_1_simple_net}, but with catalytic activity of complexes moved to $C_0$. The dashed line is the output concentration for foreign ligands in presence of self ligands ($L_{s} = 10^5, \tau_s = 0.05$ s). The effect of self ligands should be compared with Fig.~3~(d) from the paper. }
\end{center}
\end{figure}

Interestingly, keeping all kinetic parameter constant, it is possible to investigate the effect of moving all catalytic activity to the first complex in the cascade, $C_0$. The output ligand relationship is shown in Fig.~\ref{fig:par_ex_1_ol_m0}, where the dashed line is the output concentration in presence of self ligands. The output/ligand relationship does change, but without the presence of self (full lines), this network would still be regarded as achieving proper discrimination. This is not so in presence of self ligands.

As expected from our analytical analysis of the general case (Sec.~\ref{sec:general_calculation}), moving all the catalytic activity from complexes to $C_0$ leads to a catastrophic decrease in the sensitivity. The self ligands completely inhibit response at low foreign ligand concentrations.

\subsubsection{Example 2 - Non-specific Kinase Leading to Non-Monotonic Response}

\begin{figure}[h!]
\begin{center}
\includegraphics[width=0.6\textwidth]{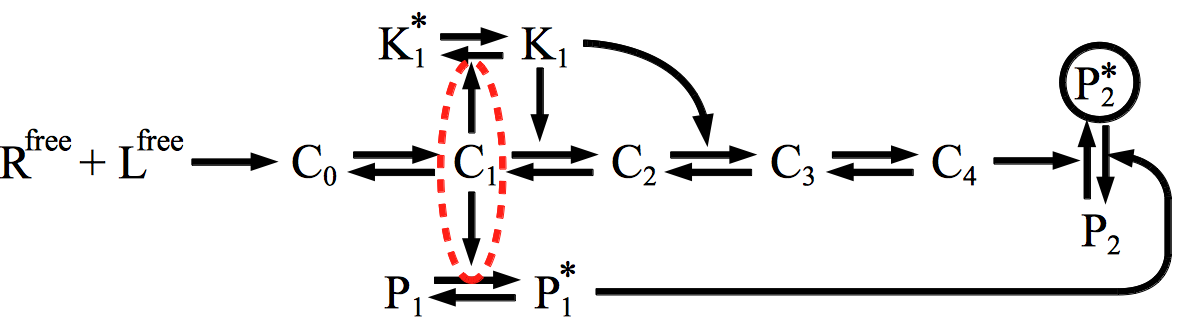}
\caption{\label{fig:par_ex_2_simple_net} Obtained network displaying parallel triage properties. Complexes $C_i$ are understood to decay to $R$ and $L$ with rate $\tau^{-1}$.}
\end{center}
\end{figure}

We display another solution has interesting features in spite its imperfection. The simplified network is shown in Fig.~\ref{fig:par_ex_2_simple_net}.

\begin{figure}[h!]
\begin{center}
\includegraphics[width=0.65\textwidth]{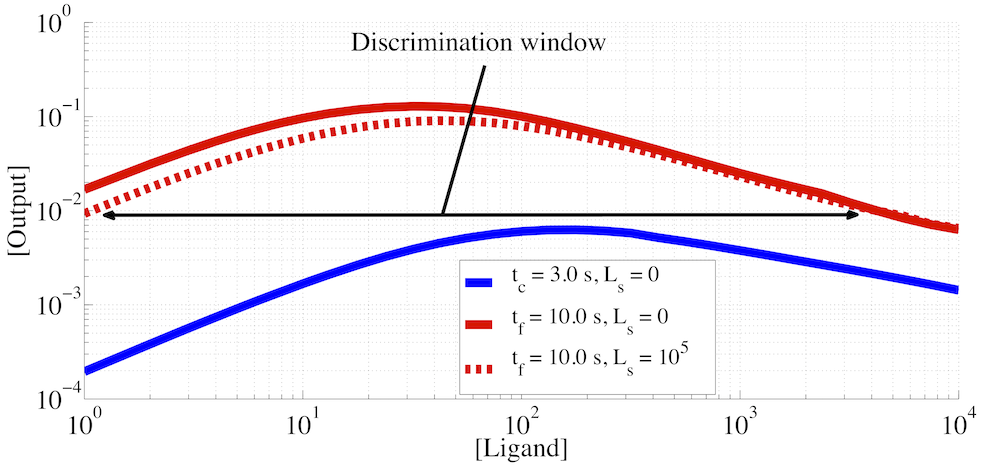}
\caption{\label{fig:par_ex_2_ol} Output versus ligand relationship (steady-state) for network of Fig.~\ref{fig:det_ex_2_simple_net}. The dashed line is the output for foreign ligands in the presence of self ligands. Notice again the little effect of self ligands in spite of their being very numerous ($10^5$).}
\end{center}
\end{figure}

As before, the complex with catalytic activity is $C_1$ (dashed circles). It is interesting to note that here, the adaptive kinase ($K_1$) is non-specific, phosphorylating two complexes in the cascade ($C_1$ and $C_2$). Since $K_1 \sim L^{-1}$, we have that [Output] $ \lesssim (K_1)^2 L \sim L^{-1}$. The output must ultimately decrease at large ligand concentration. This is indeed what is seen in the output/ligand relationship displayed in Fig.~\ref{fig:par_ex_2_ol}. Interestingly, such non-monotonic behaviour in ligand concentration (loss of response at high ligand concentration) is seen experimentally in the immune system \cite{Francois:2013,AltanBonnet:2005} We believe it to be a signature of adaptive sorting for systems with enzymes lacking the biochemical specificity needed to act on a single step in the cascade.

Here again, the adaptive sorting module is identifiable, but the non-specificity of the kinase leads to a non-monotonic response instead of a perfectly adaptive response. This non-monotonicity is sufficient to separate the output concentrations from the two ligand types over a wide range of ligand concentration.

\begin{figure}[h!]
\begin{center}
\includegraphics[width=0.65\textwidth]{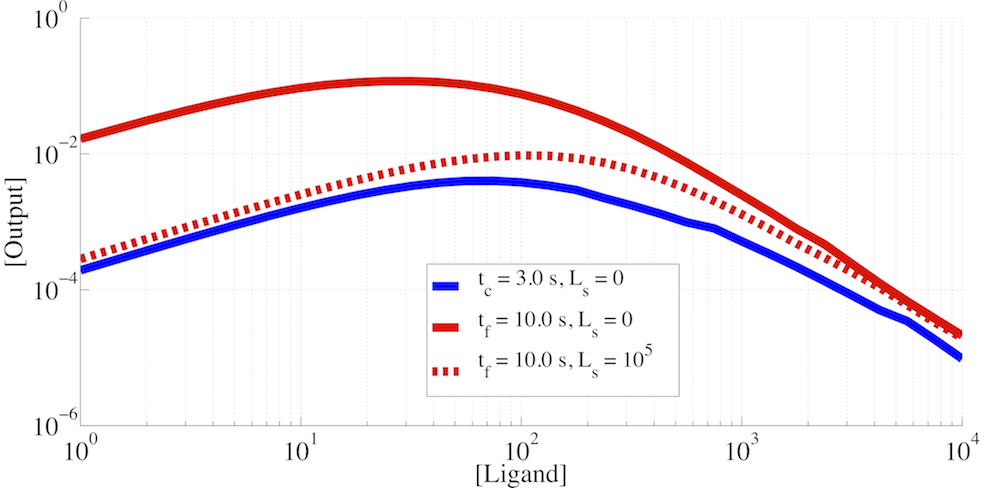}
\caption{\label{fig:par_ex_2_ol_m0} Output versus ligand relationship (steady-state) for network of Fig.~\ref{fig:det_ex_2_simple_net} with catalytic activity moved to $C_0$. The dashed line is the output for foreign ligands in the presence of self ligands. The response for foreign ligands in presence of self ligands (dashed red) is now difficultly distinguishable from that of critical non-agonist (blue) in absence of self ligands.}
\end{center}
\end{figure}

As before, for purposes of illustration, it is possible to keep the kinetic rate constants of that network fixed but move the catalytic (adaptive) activity from $C_1$ to $C_0$ and see the detrimental effect of self-ligands on the sensitivity of response. This is shown in Fig.~\ref{fig:par_ex_2_ol_m0}. The presence of a large quantity of spurious (non-specific) molecules leads a strong degradation of response if the adaptive module is activated too early in the cascade.

\clearpage

\section{Appendix -- Details for Network of Sec.~\ref{sec:phosphatase_activation}}

\begin{figure}[h!]
\begin{center}
\includegraphics[width=0.75\textwidth]{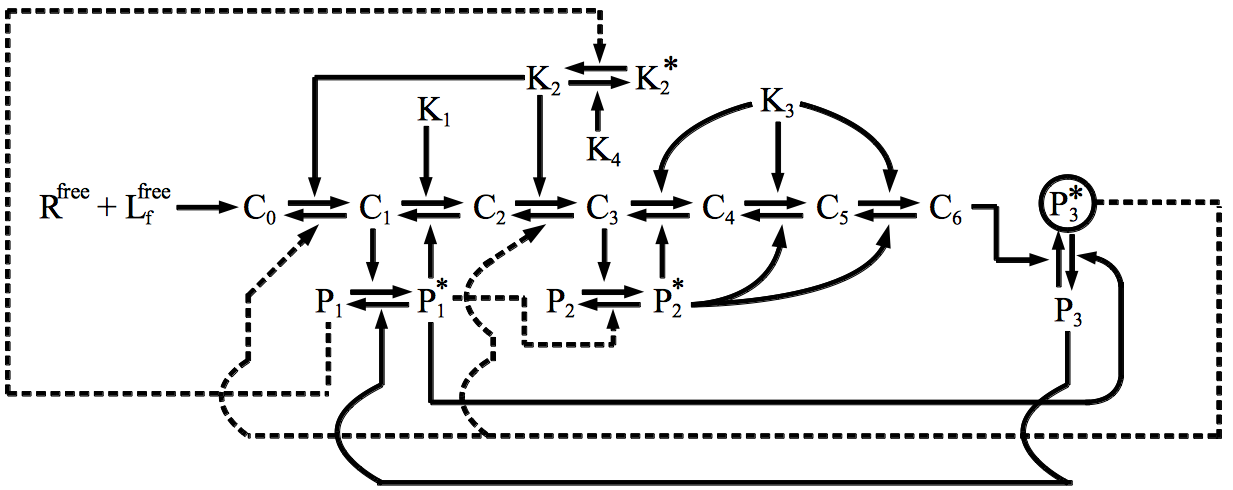}
\caption{\label{fig:par_ex_1_full} Full network as obtained from evolutionary simulations. The reactions denoted by dashed lines were not to be important for the behaviour of the system (i.e. enzymes replaced by unregulated ones without effect). All complexes $C_i$ decay to the receptor and ligands with rate $\tau_f^{-1}$ (not shown). Complexes from self are not shown for clarity.}
\end{center}
\end{figure}

We provide the details concerning the network of Sec.~\ref{sec:phosphatase_activation}, whose output concentration versus ligand concentration appears in the paper's Fig.~3~(d). Fig.~\ref{fig:par_ex_1_full} shows the actual network as obtained from evolutionary simulations. One recovers Fig.~\ref{fig:par_ex_1_simple_net} if the dashed lines are not shown. These reactions do not have to be regulated (i.e. catalyzed by enzyme which are modified by presence or absence of ligands) for the network to show its distinctive features. 

Initial concentrations and reactions rates follow in tables below. Note that the on-rate is equal to $\kappa = 7.02 \times 10^{-6} $ in the present network. All complexes $C_i$ decay to $R^{\scriptsize{\mbox{free}}}$ and $L_f^{\scriptsize{\mbox{free}}}$ with rate $\tau_f^{-1}$. All complexes $D_i$ decay to $R^{\scriptsize{\mbox{free}}}$ and $L_s^{\scriptsize{\mbox{free}}}$ with rate $\tau_s^{-1}$. 

\clearpage

\begin{scriptsize}

\begin{center}
Initial Concentrations
\end{center}

\begin{center}
\begin{tabular}{c|c||c|c}
Chemical Species & Initial Concentration & Chemical Species  & Initial Concentration\\
\hline
\hline
$K_1$ & 576 & $L_f^{\tiny{\mbox{free}}}$ & Variable\\
$K_2$ & 475& $C_0$  & 0\\
$K_2^*$ & 0 & $C_1$  & 0\\
$K_3$ &  941& $C_2$  & 0\\
$K_4$ &  861& $C_3$ & 0\\
$K_5$ &  348 & $C_4$ & 0\\
$P_1$ &  439 & $C_5$  & 0\\
$P_1^*$  & 0& $C_6$  & 0\\
$P_2$ & 717 & $L_s^{\tiny{\mbox{free}}}$ & $10^5$\\
$P_2^*$  & 0& $D_0$  & 0\\
$P_3$  & 957 & $D_1$  & 0\\
$P_3^*$  & 0& $D_2$  & 0\\
$P_4$  & 906& $D_3$  & 0\\
$R^{\tiny{\mbox{free}}}$  & $3\times 10^4$ & $D_4$  & 0\\
& & $D_5$  & 0\\
& & $D_6$  & 0\\
\end{tabular}
\end{center}

\clearpage

\begin{center}
Phosphorylations
\end{center}

\begin{center}
\begin{tabular}{c|c|c|c}
Kinase & Species & Species Phosphorylated & Rate $(\times 10^{-4})$\\
\hline
\hline
$K_1$ & $C_1$ & $C_2$ & 1.32 \\
$K_1$ & $D_1$ & $D_2$ & 1.32 \\
$K_2$ & $C_2$ & $C_3$ & 0.25\\
$K_2$ & $D_2$ & $D_3$ & 0.25\\
$C_1$ & $P_1$ & $P_1^*$ & 8.38\\
$D_1$ & $P_1$ & $P_1^*$ & 8.38\\
$K_3$ & $C_3$ & $C_4$ & 5.88\\
$K_3$ & $D_3$ & $D_4$ & 5.88\\
$K_3$ & $C_4$ & $C_5$ & 4.18\\
$K_3$ & $D_4$ & $D_5$ & 4.18\\
$K_3$ & $C_5$ & $C_6$ & 7.00\\
$K_3$ & $D_5$ & $D_6$ & 7.00\\
$C_3$ & $P_2$ & $P_2^*$ & 4.55\\
$D_3$ & $P_2$ & $P_2^*$ & 4.55\\
$K_2$ & $C_0$ & $C_1$ & 9.11 \\
$K_2$ & $D_0$ & $D_1$ & 9.11 \\
$C_6$ & $P_3$ & $P_3^*$ & 9.40\\
$D_6$ & $P_3$ & $P_3^*$ & 9.40\\
$K_4$ & $K_2$ & $K_2^*$ & 3.57\\
\end{tabular}
\end{center}

\begin{center}
Dephosphorylations
\end{center}
\begin{center}
\begin{tabular}{c|c|c|c}
Phosphatase & Species & Species Dephosphorylated & Rate $(\times 10^{-4})$\\
\hline
\hline
$P_1$ & $K_2^*$ & $K_2$ & 2.77\\
$P_3^*$ & $C_1$ & $C_0$ & 4.24\\
$P_3^*$ & $D_1$ & $D_0$ & 4.24\\
$P_1^*$ & $P_3^*$ & $P_3$ & 8.11\\
$P_3$ & $P_1^*$ & $P_1$ & 1.23\\
$P_1^*$ & $C_2$ & $C_1$ & 8.43\\
$P_1^*$ & $D_2$ & $D_1$ & 8.43\\
$P_2^*$ & $C_4$ & $C_3$ & 8.73\\
$P_2^*$ & $D_4$ & $D_3$ & 8.73\\
$P_2^*$ & $C_5$ & $C_4$ & 3.34\\
$P_2^*$ & $D_5$ & $D_4$ & 3.34\\
$P_2^*$ & $C_6$ & $C_5$ & 2.61\\
$P_2^*$ & $D_6$ & $D_5$ & 2.61\\
$P_3^*$ & $C_3$ & $C_2$ & 4.51\\
$P_3^*$ & $D_3$ & $D_2$ & 4.51\\
$P_1^*$ & $P_2^*$ & $P_2$ & 1.21\\
\end{tabular}
\end{center}
\end{scriptsize}
\clearpage

\bibliography{PRL2}

\end{document}